%% file: paper.tex
\documentclass[twocolumn]{aastex631}

\usepackage{amsmath}
\usepackage{commath}
\usepackage{xcolor}
\usepackage{CJK}

\def\siiv     {\ensuremath{\text{Si\,\textsc{iv}}}}

\def\civ     {\ensuremath{\text{C\,\textsc{iv}}}}

\def\mgii     {\ensuremath{\text{Mg\,\textsc{ii}}}}

\def\cii     {\ensuremath{\text{[C\,\textsc{ii]}}}}

\newcommand{\lya}{\ensuremath{\text{Ly}\alpha}}
\newcommand{\nv}{\ensuremath{\text{N}\,\textsc{v}}}
\newcommand{\lyaew}{\mathrm{EW}(\lya + \nv)}
\newcommand{\hi}{\ensuremath{\text{H}\,\textsc{i}}}

\def\lnu{{\rm\,erg\,s^{-1}\,Hz^{-1}}}

\newcommand{\gps}{\ensuremath{g_{\rm P1}}}
\newcommand{\rps}{\ensuremath{r_{\rm P1}}}
\newcommand{\ips}{\ensuremath{i_{\rm P1}}}
\newcommand{\ipsl}{\ensuremath{i_{\rm P1,lim}}}

\newcommand{\zps}{\ensuremath{z_{\rm P1}}}

\newcommand{\zext}{\ensuremath{z_{\rm P1,ext}}}

\newcommand{\yps}{\ensuremath{y_{\rm P1}}}

\newcommand{\yext}{\ensuremath{y_{\rm P1,ext}}}

\newcommand{\grizy}{\gps\rps\ips\zps\yps}
\newcommand{\PS}{\protect \hbox {Pan-STARRS1}}

\newcommand{\gde}{\ensuremath{g_{\rm DE}}}
\newcommand{\rde}{\ensuremath{r_{\rm DE}}}

\newcommand{\zde}{\ensuremath{Z_{\rm DE}}}

\newcommand{\Jgrond}{\ensuremath{J_{\rm G}}}
\newcommand{\Hgrond}{\ensuremath{H_{\rm G}}}

\newcommand{\intt}{\ensuremath{I_{\rm E}}}
\newcommand{\zntt}{\ensuremath{Z_{\rm E}}}
\newcommand{\Jntt}{\ensuremath{J_{\rm S}}}
\newcommand{\Hntt}{\ensuremath{H_{\rm S}}}
\newcommand{\Kntt}{\ensuremath{K_{\rm S}}}

\newcommand{\zvik}{\ensuremath{Z_{\rm VI}}}

\newcommand{\ydp}{\ensuremath{Y_{\rm R}}}

\newcommand{\slopelofarfirst}{\ensuremath{\alpha^{144\,\mathrm{MHz}}_{1.4\,\mathrm{GHz}}}}
\newcommand{\slopefirstvlass}{\ensuremath{\alpha^{1.4\,\mathrm{GHz}}_{3.0\,\mathrm{GHz}}}}
\newcommand{\sloperacsvlass}{\ensuremath{\alpha^{0.9\,\mathrm{GHz}}_{3.0\,\mathrm{GHz}}}}
\newcommand{\sloperacsfirst}{\ensuremath{\alpha^{0.9\,\mathrm{GHz}}_{1.4\,\mathrm{GHz}}}}

\def\nqso{55}
\def\nqsops{48}
\def\nqsovk{7}
\def\nradio{5} 
\def\nqlf{31}

\shorttitle{Discovery od \nqso{}  $z\approx6$ quasars}
\shortauthors{Ba\~nados et al.}
\graphicspath{{./}{figures/}}

\begin{document}
\begin{CJK*}{UTF8}{gbsn}
\title{The Pan-STARRS1 $\mathbf{\lowercase{\text{\em z}}>5.6}$ quasar survey II: Discovery of \nqso{} Quasars at $\mathbf{5.6<\text{\em z}<6.5}$}

\correspondingauthor{Eduardo Ba\~nados}
\email{banados@mpia.de}

\author[0000-0002-2931-7824]{Eduardo Ba{\~n}ados}
\affiliation{Max Planck Institut f\"ur Astronomie, K\"onigstuhl 17, D-69117, Heidelberg, Germany}
\affiliation{The Observatories of the Carnegie Institution for Science, 813 Santa Barbara Street, Pasadena, CA 91101, USA}

\author[0000-0002-4544-8242]{Jan-Torge Schindler}
\affiliation{Max Planck Institut f\"ur Astronomie, K\"onigstuhl 17, D-69117, Heidelberg, Germany}
\affiliation{Leiden Observatory, Leiden University, Niels Bohrweg 2, 2333 CA Leiden, Netherlands}

\author[0000-0001-9024-8322]{Bram P. Venemans}
\affiliation{Leiden Observatory, Leiden University, Niels Bohrweg 2, 2333 CA Leiden, Netherlands}

\author[0000-0002-7898-7664]{Thomas Connor}
\affiliation{Jet Propulsion Laboratory, California Institute of Technology, 4800 Oak Grove Drive, Pasadena, CA 91109, USA}
\affiliation{The Observatories of the Carnegie Institution for Science, 813 Santa Barbara Street, Pasadena, CA 91101, USA}
\affiliation{Center for Astrophysics $\vert$\ Harvard\ \&\ Smithsonian, 60 Garden St., Cambridge, MA 02138, USA}

\author[0000-0002-2662-8803]{Roberto Decarli}
\affiliation{INAF --- Osservatorio di Astrofisica e Scienza dello Spazio, via Gobetti 93/3, I-40129, Bologna, Italy}

\author[0000-0002-6822-2254]{Emanuele Paolo Farina}
\affiliation{Gemini Observatory, NSF's NOIRLab, 670 N A\'ohoku Place, Hilo, Hawai\'i 96720, USA}

\author[0000-0002-5941-5214]{Chiara Mazzucchelli}
\affiliation{N\'ucleo de Astronom\'ia, Facultad de Ingenier\'a y Ciencias, Universidad Diego Portales, Av.\ Ej\'ercito 441, Santiago, 8320000, Chile}

\author[0000-0001-5492-4522]{Romain A. Meyer}
\affiliation{Max Planck Institut f\"ur Astronomie, K\"onigstuhl 17, D-69117, Heidelberg, Germany}

\author[0000-0003-2686-9241]{Daniel Stern}
\affiliation{Jet Propulsion Laboratory, California Institute of Technology, 4800 Oak Grove Drive, Pasadena, CA 91109, USA}

\author[0000-0003-4793-7880]{Fabian Walter}
\affiliation{Max Planck Institut f\"ur Astronomie, K\"onigstuhl 17, D-69117, Heidelberg, Germany}

\author[0000-0003-3310-0131]{Xiaohui Fan}
\affiliation{Steward Observatory, University of Arizona, 933 N Cherry Ave, Tucson, AZ 85721, USA}

\author[0000-0002-7054-4332]{Joseph F. Hennawi}
\affiliation{Department of Physics, University of California, Santa Barbara, CA 93106, USA}
\affiliation{Leiden Observatory, Leiden University, Niels Bohrweg 2, 2333 CA Leiden, Netherlands}

\author[0000-0002-7220-397X]{Yana Khusanova}
\affiliation{Max Planck Institut f\"ur Astronomie, K\"onigstuhl 17, D-69117, Heidelberg, Germany}

\author[0000-0003-2535-3091]{Nidia Morrell}
\affiliation{Las Campanas Observatory, Carnegie Observatories, Casilla 601, La Serena, Chile}

\author[0000-0002-2579-4789]{Riccardo Nanni}
\affiliation{Leiden Observatory, Leiden University, Niels Bohrweg 2, 2333 CA Leiden, Netherlands}

\author{Ga\"el Noirot}
\affiliation{Jet Propulsion Laboratory, California Institute of Technology, 4800 Oak Grove Drive, Pasadena, CA 91109, USA}
\affiliation{Department of Astronomy \& Physics, Saint Mary's University, 923 Robie Street, Halifax, NS B3H 3C3, Canada}

\author[0000-0001-9815-4953]{Antonio Pensabene}
\affiliation{Dipartimento di Fisica ``G. Occhialini'', Universit\`a degli Studi di Milano-Bicocca, Piazza della Scienza 3, I-20126, Milano, Italy}

\author[0000-0003-4996-9069]{Hans-Walter Rix}
\affiliation{Max Planck Institut f\"ur Astronomie, K\"onigstuhl 17, D-69117, Heidelberg, Germany}

\author[0000-0003-1407-6607]{Joseph Simon}
\affiliation{Jet Propulsion Laboratory, California Institute of Technology, 4800 Oak Grove Drive, Pasadena, CA 91109, USA}
\affiliation{Department of Astrophysical and Planetary Sciences, University of Colorado, Boulder, CO 80309, USA}

\author[0000-0001-5803-2580]{Gijs A. Verdoes Kleijn}
\affiliation{Kapteyn Astronomical Institute University of Groningen}

\author[0000-0002-0125-6679]{Zhang-Liang Xie (谢彰亮)}
\affiliation{Max Planck Institut f\"ur Astronomie, K\"onigstuhl 17, D-69117, Heidelberg, Germany}

\author[0000-0002-6769-0910]{Da-Ming Yang (羊达明)}
\affiliation{Leiden Observatory, Leiden University, Niels Bohrweg 2, 2333 CA Leiden, Netherlands}

\author[0000-0002-4886-8664]{Andrew Connor}
\affiliation{Centre for Ancient Cultures, Monash University, Clayton, Vic 3800, Australia}



\begin{abstract}
The identification of bright quasars at $z\gtrsim 6$ enables detailed studies of supermassive black holes, massive galaxies, structure formation, and the state of the intergalactic medium within the first billion years after the Big Bang.
We present the spectroscopic confirmation of \nqso{} quasars at redshifts $5.6<z<6.5$ and UV magnitudes $-24.5<M_{1450}<-28.5$ identified in the optical Pan-STARRS1 and near-IR VIKING surveys (48 and 7, respectively). Five of these quasars have been independently discovered in other studies.
The quasar sample shows an extensive range of physical properties, including 17 objects with weak emission lines, ten broad absorption line quasars, and five with strong radio emission (radio-loud quasars). There are also a few notable sources in the sample, including a blazar candidate at $z=6.23$, a likely gravitationally lensed quasar at $z=6.41$, and a $z=5.84$ quasar in the outskirts of the nearby ($D\sim3\,$Mpc) spiral galaxy M81.
The blazar candidate remains undetected in NOEMA observations of the \cii\ and underlying emission, implying a star-formation rate $<30-70\,M_\odot$\,yr$^{-1}$.
A significant fraction of the quasars presented here lies at the foundation of the first measurement of the $z\sim 6$ quasar luminosity function from Pan-STARRS1 (introduced in a companion paper). The quasars presented here will enable further studies of the high-redshift quasar population with current and future facilities.
\end{abstract}



\section{Introduction} \label{sec:intro}

As the most luminous non-transient sources in the Universe, quasars enable the study of structure formation, supermassive black holes, massive galaxies, and the intergalactic medium---at unrivaled detail---when the Universe was less than a billion years old (or redshifts $z>5.6$).

The advent of various multi-wavelength large sky surveys led to a drastic increase in the number of $z>5.6$ quasars in the last decade, both at the bright ($M_{1450} \lesssim  -25$) and at the faint end ($M_{1450} \gtrsim -25$). Some of the main contributors are the Panoramic Survey Telescope and Rapid Response System (\PS; \citealt{chambers2016}), the DESI Legacy Imaging Surveys (DELS; \citealt{dey2019}), and the Dark Energy Survey (DES; \citealt{abbott2018,abbott2021}) at the bright end (e.g., \citealt{banados2016, Reed2017,Wang2019ApJ...884...30W}), while at the faint end, most quasars come from the Subaru High-z Exploration of Low-Luminosity Quasars survey (SHELLQs; e.g., \citealt{Matsuoka2018ApJS..237....5M,matsuoka2022}).

This paper presents the discovery of \nqso{}  bright quasars at $5.6<z<6.5$. Most of them were selected from \PS\ following the methods outlined in detail in \cite{banados2016}. Seven of the  quasars were selected from the VISTA Kilo-Degree Infrared Galaxy survey (VIKING; \citealt{edge2013Msngr.154...32E}), following the strategy presented by \cite{venemans2015b}.

The quasars presented in \cite{banados2016} nearly doubled the number $z>5.6$ quasars known at the time, enabling a transition from studies of individual sources to statistical analyses of the quasar population in early cosmic times.
For example, the much-enlarged quasar sample enabled
(i) the measurement of nuclear chemical enrichment, black hole mass, and Eddington ratio distributions (e.g., \citealt{schindler2020,farina2022,lai2022, wang2022ApJ...925..121W});
(ii) the characterization of their X-ray (e.g., \citealt{vito2019A&A...630A.118V}) and radio (e.g., \citealt{liu2021}) properties;
(iii) the search for signatures of outflows and black hole feedback (e.g., \citealt{meyer2019,novak2020, bischetti2022});
(iv) a census of (atomic/molecular) gas and dust in the quasar hosts (e.g., \citealt{decarli2018,decarli2022, venemans2018,li2020,pensabene2021}), including the serendipitous discovery of star-forming companion galaxies \citep{decarli2017};
(v) the search for extended \lya\ nebular emission (e.g., \citealt{farina2019}) and its connection to \cii\ gas \citep{drake2022};
(vi) the quantification of the properties of water reservoirs in these quasars (e.g., \citealt{Pensabene2022});
(vii) the identification of a population of particularly young quasars \citep{eilers2020};
(viii) the first constraints on quasar clustering at $z\sim 6$ \citep{greiner2021};
(ix) the study of the environments where these quasars reside \citep[e.g.,][]{farina2017,mazzucchelli2017a,meyer2020,meyer2022};
(x) the study of heavy elements in intervening absorption systems at $z>5$ (e.g., \citealt{chen2017,cooper2019});
(xi) quantitative constraints on the thermal state of the intergalactic medium at $z>5$ (e.g., \citealt{gaikwad2020}); and
(xii) constraints on the end phases of cosmic reionization (e.g., \citealt{bosman2022}).
In addition to these population studies, the large sample of quasars also enabled the identification of exciting individual sources whose properties stand out. They have been studied in much more detail in several works  (e.g., \citealt{banados2018a, banados2019b, connor2019, Connor2021ApJ...911..120C, decarli2019, momjian2018,momjian2021, rojas2021, vito2021A&A...649A.133V}).

This paper is structured as follows. Section~\ref{sec:selection} describes the main selection strategies and follow-up photometry used to identify \PS\ and VIKING quasar candidates. Section~\ref{sec:spectroscopy} presents the spectroscopic observations of the quasar discoveries and discusses some individual sources. In Section~\ref{sec:radio}, we  discuss \nradio{} radio-loud quasars found in this work. We summarize our results in Section~\ref{sec:summary}. We note that \nqlf{}  quasars presented here are part of the $z\sim 6$ \PS\ quasar luminosity function presented in a companion paper by \cite{schindler2022}. Table~\ref{tab:spectroscopy} lists the spectroscopic observations of the quasars.  Table~\ref{tab:qsos-info} reports their coordinates and general properties, and Table~\ref{tab:qsos-phot} describes their photometry.

  All magnitudes reported in the paper are in the AB system, and limits correspond to $3\sigma$.
The \PS\ magnitudes used and reported in this paper are \textit{dereddened} \citep{schlafly2011}.
We use a standard flat $\Lambda$CDM cosmology with  $H_0 = 70 \,\mbox{km\,s}^{-1}$\,Mpc$^{-1}$, $\Omega_M = 0.30$.

\section{Candidate selection}\label{sec:selection}

The main quasar selection procedures are described in detail in \cite{banados2014,banados2016} for candidates selected from \PS\ and in \cite{venemans2015b} for candidates selected with VIKING. For completeness, we briefly summarize the selection criteria below.

\subsection{\PS\ candidate selection}\label{sec:ps1_selection}
Pan-STARRS1 imaged all the sky north of declination $-30^\circ$ in five filters, $\grizy$, multiple times \citep{chambers2016}.
To select high-redshift quasar candidates, in \cite{banados2014,banados2016} we used the first and second internal releases of the stacked \PS\ catalog. This time, we used the PV3 final internal version, which is very close to the final public \PS\ release hosted by the \textit{Barbara A. Mikulski Archive for Space Telescopes} at the Space Telescope Science Institute. The most noticeable difference is the astrometric accuracy as the public version has been tied to Gaia astrometry \citep{2016A&A...595A...1G}. Here we report the PV3 catalog entries as
used in our selection but the coordinates in Table~\ref{tab:qsos-info} correspond to the  recently updated ones\footnote{see `PS1 news' in \url{https://panstarrs.stsci.edu} and \cite{lubow2021}.} on 2022 June 30 using Gaia Early Data Release 3 \citep{2021A&A...649A...1G}.

In our search we exclude the region near M31 ($7\degr< \,$R.A.$ <14\degr$; $37\degr< \,$Decl.$ <43\degr$) and the most dust obscured regions in the Milky Way, requiring a reddening of $E(B-V)<0.3$ \citep{schlegel1998}. This allows to find quasars in less obscured regions of the Galactic plane (even at Galactic latitude $\left| b \right| < 20\degr$). However, the number density of sources at such low Galactic latitudes makes it harder to efficiently find quasars. In this work we present only one new quasar with $\left| b \right| < 20\degr$: P119+02 at $b=15.48\degr$.

We required that at least 85\% of the expected point-spread function (PSF)-weighted flux in the \ips, \zps\, and \yps\ resides in valid pixels. We exclude sources whose measurements in the  \zps\ and \yps\ are suspicious according to the Imaging Processing Pipeline \citep{Magnier2020PixelAnalysis}, using the same quality flags described in Table 6 of \cite{banados2014}. We required our candidates to be point sources using ($-0.3<\zext < 0.3$) \textit{or}   ($-0.3< \yext <0.3$), where   \mbox{mag}\ensuremath{_{\rm P1,ext}} is  the difference between the aperture and PSF magnitudes. This removes $\sim$92\% of galaxies while recovering $\sim$93\% of stars and $\sim$97\% of quasars (see Section~2.1 in \citealt{banados2016}).

To select high-redshift quasar candidates we applied the following color and S/N requirements:

\begin{subequations}
\begin{eqnarray}
 (\mbox{S/N}(\gps) < 3)  & \nonumber  \\
((\mbox{S/N}(\ips) \geq 3 )\; \mbox{AND} \, (\ips - \zps > 2.0)) \; \mbox{OR} & \nonumber \\
 (\ipsl - \zps > 2.0) & \nonumber \\
 \nonumber
\end{eqnarray}
\end{subequations}

The locations of the discovered quasars in the $\ips-\zps$ vs.\ $\zps-\yps$ plane are shown in the left panel of Figure~\ref{fig:ps1-selection}.
We have additional criteria that are different if the candidates have a  $\zps-\yps<0.5$ color  (candidates expected to be at a redshift between 5.6 and 6.2) or
$\zps-\yps \geq0.5$ (candidates expected to be at redshifts between 6.2 and 6.5). For sources with $\zps-\yps<0.5$:

\begin{subequations}
\begin{eqnarray}
\mbox{S/N}(\zps) >  10& \nonumber \\
\mbox{S/N}(\yps) > 5 & \nonumber \\
\mbox{S/N}(\rps) < 3 \; \mbox{OR}  \; (\rps - \zps > 2.2) & \nonumber 
\end{eqnarray}
\end{subequations}

And for sources with $\zps-\yps\geq0.5$:

\begin{subequations}
\begin{eqnarray}
\mbox{S/N}(\zps) >  7& \nonumber \\
\mbox{S/N}(\yps) > 7 & \nonumber \\
\mbox{S/N}(\rps) < 3  \nonumber 
\end{eqnarray}
\end{subequations}

We then performed our own forced photometry (using an aperture that maximizes the S/N of stars in the field) in both the stacked and single epoch \PS\ images to ensure the measured colors are consistent with the catalogued ones and to remove spurious or moving objects (see Sections 2.2 and 2.3 in \citealt{banados2014}). There is no human intervention in the previous steps but we finally visually inspect the stacked and single-epoch images to remove remaining obvious bad candidates.

We note that the $z\sim 6$ quasar luminosity function presented in \cite{schindler2022} focuses on objects with $\zps-\yps<0.5$ for which our spectroscopic follow-up has been more extensive.

\subsection{VIKING candidate selection}\label{sec:viking_selection}

The VIKING survey covers 1350\,$\deg^2$ of the sky in five bands $\zvik YJHK_S$. The first requirement for our quasar selection is that the candidates need to be detected in at least the \zvik\ and $Y$ bands but faint or not detected in the Kilo-Degree Survey (KiDS). KiDS is a public survey that covers almost the same area as VIKING but in the optical Sloan $ugri$ bands \citep{kuijken2019}. We then retained sources  classified as point sources in the VIKING catalog (with a galaxy probability $P_{\rm gal}< 0.95$, see \citealt{venemans2013} for details). To select quasar candidates in the redshift range $5.8\lesssim z \lesssim 6.4$, we adopt the following criteria:

\begin{subequations}
\begin{eqnarray}
\mbox{S/N}(\zvik) > 7 \nonumber \\
 i- \zvik > 2.2 \nonumber \\
 \zvik - Y < 1.0 \nonumber \\
  -0.5 < Y -J < 0.5 \nonumber
\end{eqnarray}
\end{subequations}

\begin{figure*}[ht]
\centering
\includegraphics[scale=0.65]{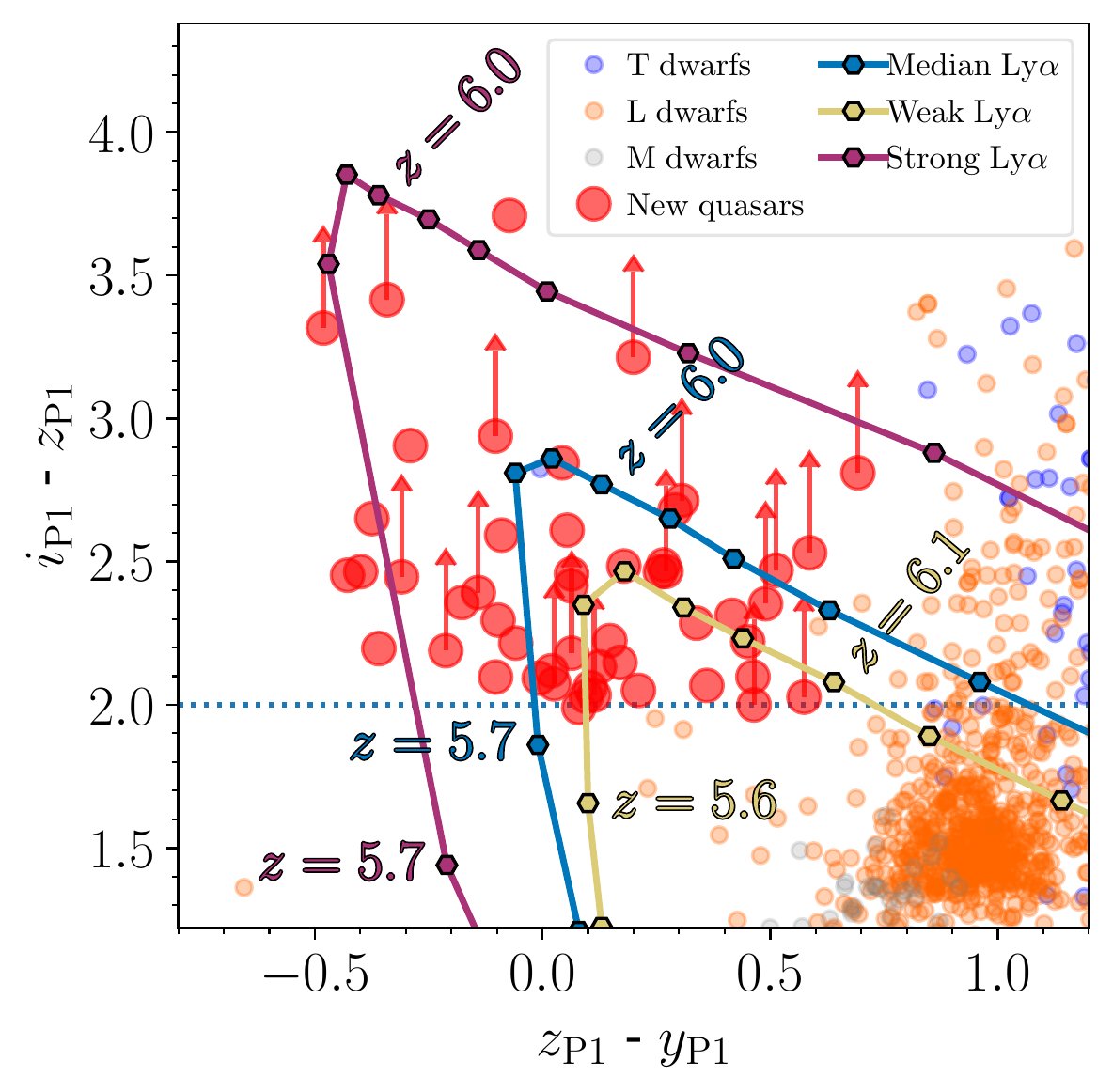}
\includegraphics[scale=0.65]{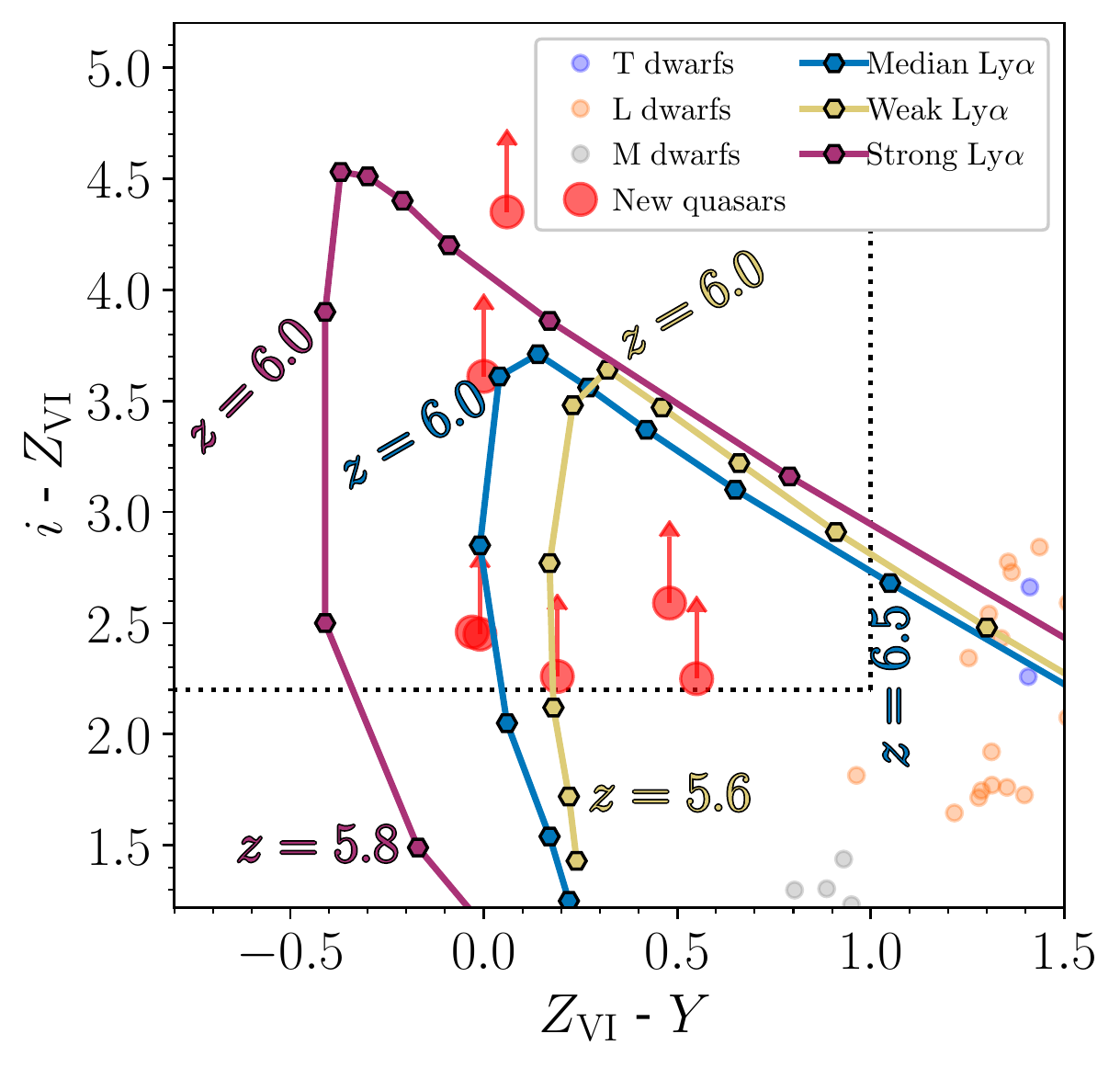}
\caption{
\textit{Left: }$\ips-\zps$ vs.\ $\zps - \yps$ diagram showing the main \PS\ color criteria, the location of the discovered quasars (red circles), and the compilation of $MLT$ dwarfs from \cite{banados2016}. The solid lines represent the color tracks of the $z\sim$ 6 quasar composites with `strong', `median', and `weak' \lya\ line from \cite{banados2016}. The symbols are plotted in steps of $\Delta z = 0.1$, and two redshifts are labeled for each track.
Note that the four VIKING quasars without PS1 information are not shown here (see Table \ref{tab:qsos-phot}).
\textit{Right: } Same as for the left panel but for $i-\zvik$ vs.\ $\zvik - Y$ diagram showing the main KiDS-VIKING color criteria and the location of the discovered quasars.
}
\label{fig:ps1-selection}
 \end{figure*}

We performed forced photometry on the KiDS and VIKING pixel data to verify the catalog magnitudes and non-detections.
All our candidates are visually inspected and we use the fact that VIKING data are usually taken in different nights, including two epochs in $J$, to remove moving and variable objects.
The locations of our discovered quasars in the  $i_{\rm forced} - \zvik$ vs.\  $\zvik-Y$ plane are shown in the right panel of Figure~\ref{fig:ps1-selection}.

\subsection{Follow-up and survey photometry} \label{sec:follow-up-phot}

We have compiled extra information from public surveys and obtained follow-up photometry for our quasar candidates. In some cases this extra information was used to prioritize objects for spectroscopic follow-up. We also obtained follow-up photometry for quasars after their spectroscopic discovery to constrain their spectral energy distribution and secure absolute flux calibration for future near-infrared spectroscopic observations. In Table~\ref{tab:qsos-phot} we report follow-up photometry obtained in the following filters and telescopes: $I\#705$ (\intt) and $Z\#623$ (\zntt) with the ESO Faint Object Spectrograph and Camera \citep[EFOSC2;][]{snodgrass2008} at the NTT telescope in La Silla, \Jgrond\ and \Hgrond\ with the GROND camera \citep{greiner2008} at the MPG 2.2\,m telescope in La Silla, \Jntt\Hntt\Kntt\ with SOFI \citep{moorwood1998} at the NTT telescope in La Silla, and \ydp\ with the RetroCam\footnote{\url{http://www.lco.cl/?epkb_post_type_1=retrocam-specs}} camera at the Du Pont telescope in Las Campanas Observatory.

We report the photometry and observation dates in Table \ref{tab:qsos-phot}. For completeness, in Table \ref{tab:qsos-phot} we also report $IZYHK$ magnitudes from public surveys or published papers when available. The public surveys are  DELS \citep{dey2019}, DES \citep{abbott2018,abbott2021}, KiDS \citep{kuijken2019}, UHS \citep{dye2018}, UKIDSS \citep{lawrence2007}, VHS \citep{mcmahon2013}, and VIKING \citep{edge2013Msngr.154...32E} .

Finally, we note that most of our VIKING-selected candidates were followed-up with EFOSC2 observations in the \intt\ band. Sources that satisfied $\intt-\zvik>1.0$ were considered high priority for spectroscopic follow-up.

\section{Discovery of \nqso{} quasars}\label{sec:spectroscopy}

\subsection{Spectroscopic Observations} \label{sec:spectroscopy-log}

Here we present the discovery of \nqso{}  quasars. \nqsops{} selected from \PS\ and \nqsovk{} from VIKING (see Section~\ref{sec:selection}). We note that two of the three VIKING quasars with \PS\ information satisfy all of our \PS\ selection criteria except for S/N(\zps)$>10$ (8.8 and 9.0 for J0046--2837 and J2315--2856, respectively).
J2318--3029, on the other hand, would have been rejected by three criteria: low S/N(\zps)$=6.1$,  classified as extended in both \zps\ and \yps, and its \yps\ band was flagged because its moments were not measured due to low S/N (flag MOMENTS\_SN = 0x00040000).
This shows that further \PS\ quasars could be found by relaxing the criteria from Section~\ref{sec:ps1_selection}. Indeed, we did find two quasars by loosening some of the criteria. In one case the selection was enabled by requiring radio emission, as discussed in Section \ref{sec:notes_extended}.

\begin{figure*}[h!]
\centering
\includegraphics[scale=0.8]{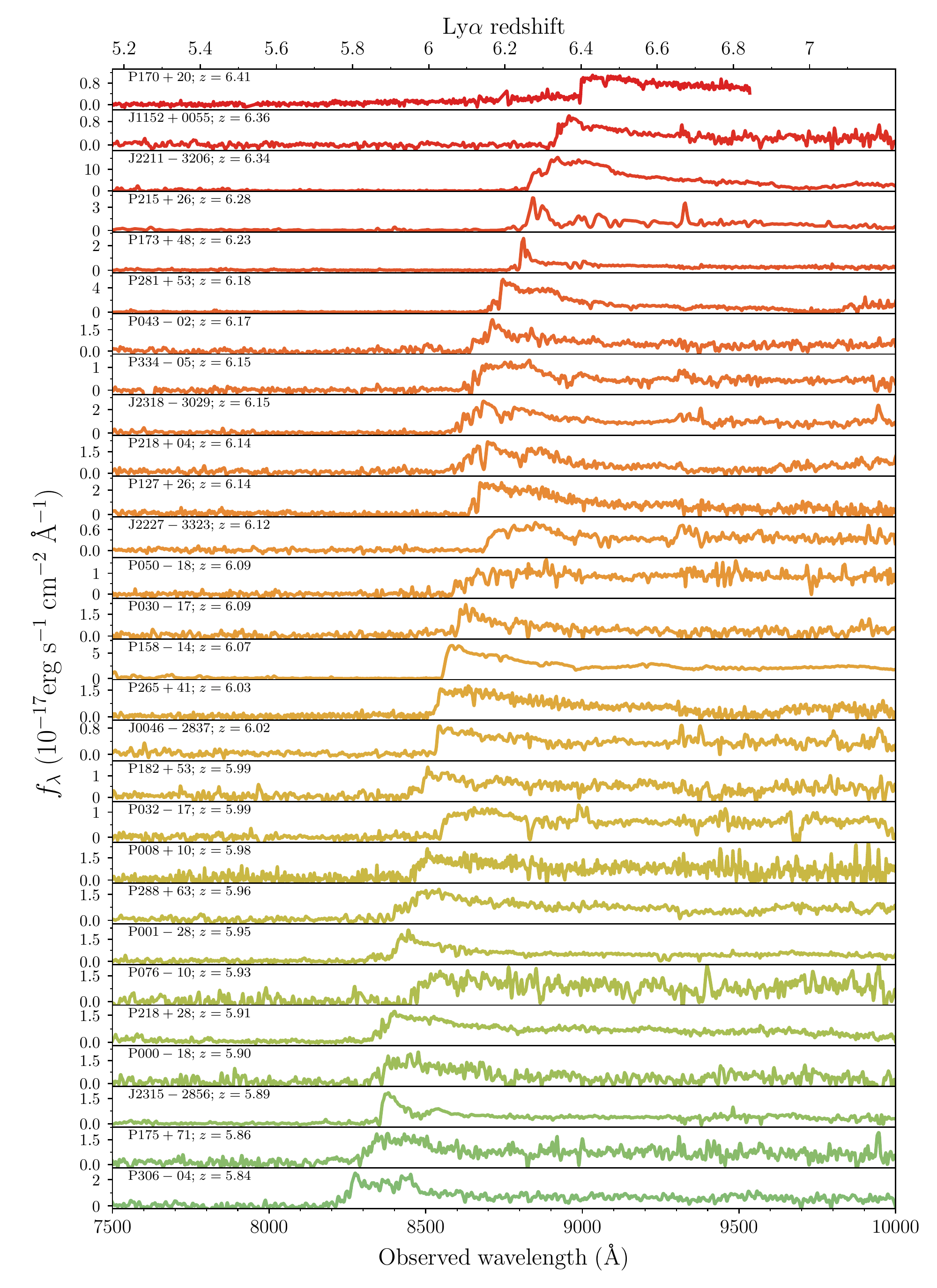}
 \caption{Discovery spectra of the \nqso{} $z\sim 6$ quasars presented in this work, sorted by decreasing redshift.
 \label{fig:qso-spectra}}
 \end{figure*}

The discovery of the \nqso{} quasars presented in this work has been a large effort, involving multiple observatories in the time frame 2013 Nov 19 - 2022 Sep 28.
Five of these quasars have been independently discovered by other groups: P158--14 by \cite{chehade2018}, P173+48 and P207+37 by \cite{gloudemans2022}, and P127+26 by Warren et al.\ in prep.
A few of these quasars have been part of multiple follow-up campaigns and some of their properties already presented in the literature (e.g., \citealt{decarli2018,eilers2020,venemans2020,bischetti2022}). These quasars are discussed in Section~\ref{sec:individual_notes}.  The spectroscopic observations log is listed in Table~\ref{tab:spectroscopy} and the main quasar properties are presented in Table~\ref{tab:qsos-info}. Quasars with multiple observations in Table~\ref{tab:spectroscopy} are due to the first spectrum being of low quality or taken under very bad weather conditions and were followed-up later on. The latest spectra are shown in Figure~\ref{fig:qso-spectra}.

The spectrographs/telescopes used for the discovery of these quasars are:
the Double Spectrograph  \citep[DBSP;][]{oke1982} on the 200-inch (5\,m) Hale telescope at Palomar Observatory (P200),
the Low Dispersion Survey Spectrograph (LDSS3; \citealt{stevenson2016A})
at the Clay Magellan telescope at Las Campanas Observatory,
the Multi-Object Double Spectrograph  \citep[MODS;][]{pogge2010} at
the Large Binocular Telescope (LBT),
EFOSC2
at the NTT telescope in La Silla,
the Low Resolution Imaging Spectrometer \citep[LRIS;][]{oke1995} at the Keck telescope on Mauna Kea,
the Gemini Multi-Object Spectrographs \citep[GMOS;][]{hook2004} on the Gemini-North telescope,
the Red Channel Spectrograph \citep{schmidt1989} on the 6.5\,m MMT Telescope, and the FOcal Reducer/low dispersion Spectrograph 2 \citep[FORS2;][]{appenzeller1992}
at the Very Large Telescope (VLT).

The spectra were reduced with standard procedures, including bias subtraction, flat fielding, sky subtraction, extraction, and wavelength and flux calibration.
All  observations taken before 2020 were reduced with the Image Reduction
and Analysis Facility \citep[IRAF;][]{tody1986}.
Observations taken after 2020 were reduced with the Python Spectroscopic Data Reduction Pipeline  \citep[PypeIt;][]{pypeit:joss_arXiv,pypeit:joss_pub,pypeit:zenodo} except for the P200/DBSP which were reduced with IRAF and NTT/EFOSC2 data which were reduced with the EsoReflex pipeline \citep{freudling2013}. The spectra were absolute flux calibrated by matching the spectra to their \zps\ magnitude (or \zvik\ when no \zps\ exists). The \nqso{} spectra sorted by descending redshift are shown in Figure~\ref{fig:qso-spectra}.

\begin{figure*}[h!]
\centering
\includegraphics[scale=0.8]{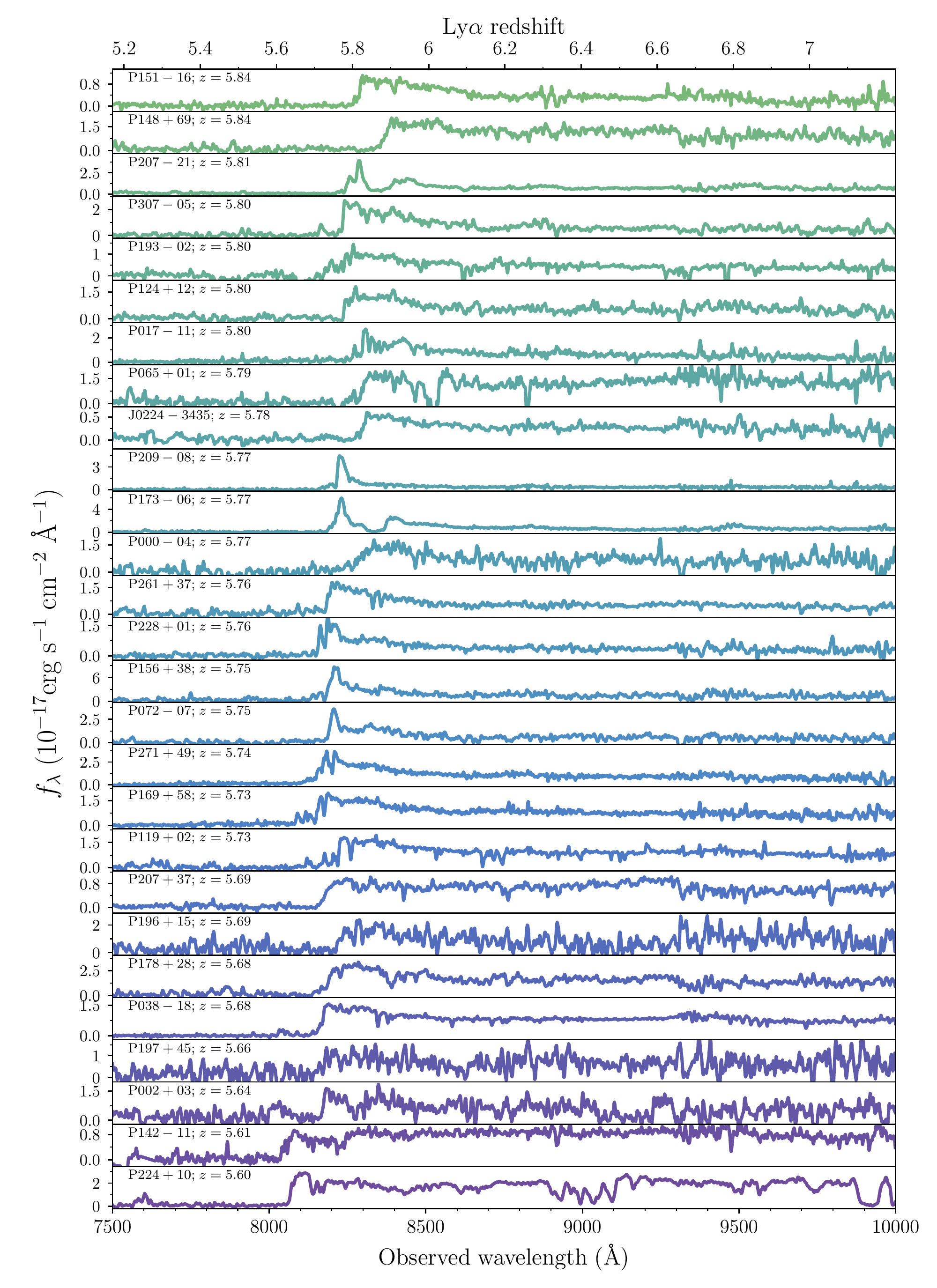}\\

{\textbf{Figure~\ref{fig:qso-spectra}}: Continuation}
 \end{figure*}

\subsection{Redshifts and continuum magnitudes} \label{sec:redshift}

Different \lya\ profiles can significantly impact
quasar colors and expected redshifts. From the colors of the  quasars shown in Figure~\ref{fig:ps1-selection}, we expect a significant fraction to have weak or `median' \lya\ and only a few with strong emission lines.

To estimate the redshifts, we fit the spectra of the quasars with six different quasar templates to encompass an extensive range of emission line properties (especially $\lya$).  The six templates are:
\begin{itemize}
    \item[(i)] \textit{vandenberk2001} that is the median of more than 2000 quasar spectra from the Sloan Digital Sky Survey \citep{vandenberk2001}.
    \item[(ii)] \textit{selsing2016} that is the median of seven bright $1<z<2$ quasars \citep{selsing2016}.
    \item[(iii)] \textit{yang2021} that is the median spectra of 38 $z>6.5$ quasars, as presented in \cite{yang2021}.
    \item[(iv)] \textit{median-\lya} that is the median of 117 $z\sim 6$ quasar from \cite{banados2016}.
    \item[(v)] \textit{strong-\lya} that is the median of the 10\% of the $z\sim 6$ spectra with the largest rest-frame $\lyaew$ from \cite{banados2016}.
    \item[(vi)]  \textit{weak-\lya} that is the median of the 10\% of the $z\sim 6$ spectra with the smallest rest-frame $\lyaew$ from \cite{banados2016}.
\end{itemize}

The \lya\ profiles of (iii) and (iv) are very similar. The templates (iv--vi) only cover up to $1400\,$\AA. We  stitch them at rest-frame wavelength $1300\,$\AA\ with the \textit{yang2021} template so that we can use the best-fitting templates to also derive rest-frame magnitudes.

After masking prominent absorption features, we choose the best-fitting template with the minimum
 $\chi^2$ in the $1212-1400\,$\AA\ wavelength range. This procedure works well for most cases but with a few exceptions.
The most common case of quasars that are not well fit are those with narrow \lya\ and \nv\ lines that are not represented in our templates. For most of the quasars that are not well fit,  the strength of the \lya\ line is between the \textit{median-\lya} templates and \textit{strong-\lya}  (quasars P000--18, P030--17, P072--07, P156+38, P173--06, P173+48, P207--21, J2315--2856). The next common cases are BAL quasars with narrow \nv\ and very absorbed \lya\ (quasars P215+26, P281+53, J2211--3206). In all the cases discussed above, the \textit{strong-\lya} template gives a clear superior fit (visually) even if it does not provide the smallest $\chi^2$. This is because the \textit{strong-\lya} template is the only one including a prominent narrow \nv\ line. We report the values derived using
the \textit{strong-\lya} template in all these cases. P170+20 is the other exception, which is discussed further in Section~\ref{sec:individual_notes}.

To quantify the uncertainty in our redshift estimates, we did the following.
 We compiled a list of published \mgii\ and \cii\ redshifts for $z\gtrsim 6$ quasars, for which we had their discovery spectra (i.e., of similar quality to the spectra analyzed in this paper). This resulted in 39 quasars with \mgii\ redshifts \citep[from][]{schindler2020,mazzucchelli2017b,onoue2019,shen2019} and 27 with \cii\ redshifts \citep[from][]{decarli2018,eilers2020,mazzucchelli2017b,meyer2022,venemans2020,wang2011,wang2013}.
 We then calculated the difference between these redshifts and the template fitting redshifts for these quasars. The mean difference and standard deviation are $-0.01 \pm 0.02$ and $+0.01 \pm 0.03$ for the $z_{\rm MgII}-z_{\rm temp}$ and $z_{\rm [CII]}-z_{\rm temp}$, respectively. Therefore, the typical uncertainties in our redshift estimates are $<0.03$.

Table~\ref{tab:qsos-info} lists the measured redshifts, rest-frame magnitudes at 1450\,\AA, and the best-fit templates used to derive the rest-frame magnitudes and, in most cases, the redshifts. If \cii-redshifts are available, we use those values \citep{decarli2018,eilers2020,venemans2020}. In one case (P173+48) we use an absorption feature to determine the redshift (see Section \ref{sec:notes_extended}).

\subsection{Notes on selected objects}
\label{sec:individual_notes}

Here we present additional notes on selected objects, sorted by R.A.

\subsubsection{VIK~J0046--2837 ($z=6.02$)}
ALMA observations for this quasar were presented in \cite{decarli2018}, finding a faint cold dust continuum detection and no \cii\ line. \cite{schindler2020} report a \mgii\ redshift of $z=5.993\pm 0.002$ and \cite{farina2022} estimate a black hole mass of $M_{\rm BH}=3.5\times 10^8\,M_\odot$.

\subsubsection{PSO~J065.9589+01.7235 ($z=5.79$)}
PSO~J065.9589+01.7235 had $\zps-\yps=0.6$ in PV2 and that is why this is not part of the luminosity function presented in \cite{schindler2022}.  However, its color
in the latest version of PS1 is $\zps-\yps=0.34$ and it would have been selected. This quasar is also part of the XQR-30 sample (D'Odorico et al.\ in prep.; \url{https://xqr30.inaf.it}) and it was classified as a broad absorption line quasar by \cite{bischetti2022}. Our template redshift of $z=5.79$ is consistent with the \mgii\ redshift $z=5.804$ reported by \cite{bischetti2022}.

\subsubsection{PSO~J127.0558+26.5654 ($z=6.14$)}
This quasar is also known as J0828+2633 (Warren et al.\ in prep.). It has been reported and studied in other published work (e.g., \citealt{mortlock2012,ross2020}) and an actual optical spectrum was first shown in \cite{li2020}. Our redshift estimate derived from template fitting ($z=6.14$) is significantly different to the redshift used in the literature of $z=6.05$, which is only possible if the $\lya$ emission was completely absorbed. For completeness, the redshift derived using our \textit{weak-\lya} template is $z=6.08$.

\subsubsection{PSO~J148.4829+69.1812 ($z=5.84$)}
This quasar is located in the outskirts of M81, and therefore myriad multi-wavelength data in the field already exist. Furthermore, its location is suitable for parallel observations with JWST covering both the quasar field and M81. The $\lya$ line is weak or heavily absorbed. The detailed spectral energy distribution and environment studies of this source will be presented in a separate paper.

\subsubsection{PSO~J158.6937--14.4210 ($z=6.0685$)}
This quasar was independently discovered by \cite{chehade2018} and it has appeared in a number of other articles (e.g., \citealt{eilers2020,schindler2020}). \cite{eilers2020} measure a very small proximity zone and classify this source as a ``young quasar''. Our best-fitting template is \textit{median-\lya} yielding $z=6.06$, in good agreement with  $z=6.0685 \pm 0.0001$ as measured from the \cii\ line from the host galaxy \citep{eilers2020}.
To measure the rest-frame magnitudes we use the \textit{median-\lya} template at the \cii\ redshift.

\subsection{PSO~J170.8326+20.2082 ($z=6.41$)}
\begin{figure*}[htb]
\centering
\includegraphics[width=\textwidth]{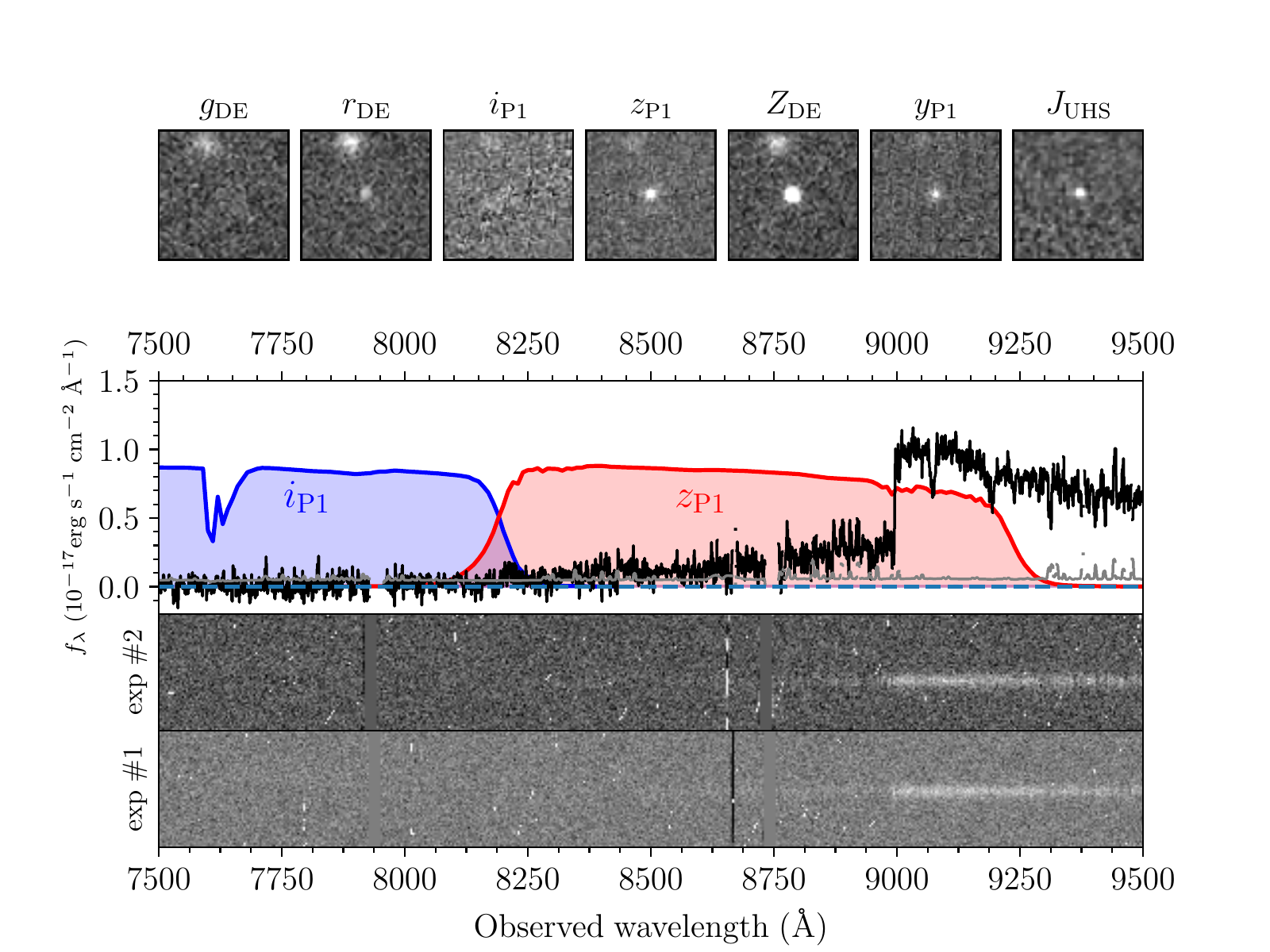}
\caption{
Gravitationally lensed quasar candidate P170+20 at $z=6.41$. The extracted 1D spectrum (shown in black and th the $1\sigma$ error vector in in gray) is the composite from the extraction of the two independent exposures shown at the bottom (sky-subtracted spectra, divided by the noise). In the 1D and 2D spectra, an excess of flux blueward of the sharp break is evident at  at $\sim$9000\,\AA. The top panel displays $15\arcsec\times 15\arcsec$ postage stamps for images in the DELS, \PS, and UHS surveys. For reference, the \ips\ and \zps\ filter curves are overplotted on top of the 1D spectrum. The \ips\ image is not reliable at the position of the quasar. The clear detection in the \rde\ image is not expected for a source at $z=6.4$ (wavelength region not covered by our spectrum).}
\label{fig:lensed}
 \end{figure*}

This quasar shows the most peculiar spectrum in the sample. We show postage stamps and 1D and 2D spectra of
this quasar in Figure~\ref{fig:lensed}.
There is an extremely sharp break in flux at $8992\,$\AA, consistent with a $z\sim 6.4$ quasar. Such a high redshift is at odds with its blue   $\zps-\yps<0$ color (see Section~\ref{sec:ps1_selection}). An excess of flux blueward of
the break explains the color. This excess declines smoothly and is inconsistent with high transmission in the \lya\ forest.

The spectrum shows no clear emission lines, and the most plausible explanation we could find  is that this source could be a lensed quasar. In this scenario, the flux blueward of the break corresponds to a foreground source. If confirmed, this would make it only the second gravitationally lensed quasar known at $z>5$  \citep[see ][]{fan2019,andika2022}. The closest visible source in the \PS\ survey is 6\arcsec\ away, too far for being the potential foreground lensed source. The extremely sharp break would imply a very small proximity zone in line with the expectations for gravitationally lensed quasars \citep{davies2020ApJ...904L..32D}. Moreover, the system is clearly detected in the DELS \rde\ image (see Fig.~\ref{fig:lensed}), with $\rde=23.11\pm0.09$ (and $\gde>25.1$ and $\zde=20.31\pm 0.02$). Detection in the $\rde$ is not expected for a source at $z=6.4$, and this emission is likely coming from the potential foreground galaxy. We will defer a more detailed analysis of the source when we have further follow-up observations. Higher resolution observations from space, Integral Field Unit observations, (sub)mm spectral line scans, and/or X-ray observations (e.g., \citealt{Connor2021ApJ...922L..24C}) are required to confirm or provide supporting evidence for the lensing scenario.

As for redshift determination, the best-fitting template is \textit{weak-\lya} at $z=6.35$ but all the other templates (except \textit{strong-\lya}, which would not apply here) yield $z=6.41$, consistent with the observed \lya\ break. Given the peculiarity of the source and the lack of clear features, we adopt $z=6.41$ and use the template \textit{yang2021} (the second best fit) to derive the rest-frame magnitudes, for which we do not attempt to correct for a possible magnification in this work.

\subsection{VIK~J1152+0055 ($z=6.3643$)}
This quasar was independently discovered by \cite{matsuoka2016} and it has been widely studied since then.
\cite{decarli2018} and \cite{izumi2018} studied the \cii\ and cold dust properties with ALMA while \cite{onoue2019} measured the black hole mass from the \mgii\ emission line ($M_{\rm BH}=6.3\times 10^8\,M_\odot$). Our best-fitting template is \textit{vandenberk2001}, yielding $z=6.36$ in agreement with the more accurate \cii-redshift $z=6.3643$ from \cite{decarli2018}.

\subsubsection{PSO~J265.9298+41.4139 ($z=6.0263$)}
P265+41 is a BAL quasar. Our best-fit template is the \textit{median-\lya}, yielding $z=6.05$, and the second best fit is $z=6.04$ with the \textit{yang2021} template. However, if we fix the redshift  derived from the \cii\ observations ($z_{\rm \cii}=6.0263\pm0.0001$, \citealt{eilers2020}), the best-fit template is \textit{yang2021} and that is what we used to derive the rest-frame magnitudes. P265+41 is the brightest quasar in \cii\ and dust continuum emission studied by \cite{eilers2020}, indicating that the quasar resides in a starburst galaxy with a star-formation rate SFR$\gtrsim1470\,M_\odot\,$yr$^{-1}$.  We note that P265+41 has one of the reddest $\yps-J$ colors. This is because the strong \siiv\ BAL falls exactly in the \yps\ filter.

\subsection{VIK~J2211--3206 ($z=6.3394$)}
This is the brightest quasar of the sample ($M_{1450}=-28$) and it has already been part of a number of studies even before its discovery publication.  J2211--3206 shows very strong BAL features studied by \cite{bischetti2022}. The \cii\ and dust from the host galaxy was reported by \cite{decarli2018} and \textit{HST} imaging of the quasar and its environment was presented in \cite{mazzucchelli2019}. Our template redshift fails at finding a good solution for a quasar with such strong BAL absorption.  \cite{bischetti2022} report a \mgii-redshift $z=6.330$ while \cite{decarli2018} report a \cii-redshift $z=6.3394$. We note that the $Y$- and $J$-bands are strongly contaminated by \siiv\ and \civ\ BAL absorption, and therefore it would be difficult to obtain the intrinsic rest-frame magnitudes from the observed spectrum.  We use our \textit{strong-\lya} template at the \cii\ redshift to estimate the rest-frame magnitudes.

\subsection{VIK~J2318--3029 ($z=6.1456$)}
Our best-fitting template is \textit{median-\lya} with $z=6.14$, in good agreement with the \cii\ redshift $z=6.1456$ reported in \cite{venemans2020}.

\subsection{Notes on objects not satisfying selection criteria}
\label{sec:notes_extended}

In this section we discuss two \PS\ quasars from an extended selection, which would have not been selected using the criteria of Section \ref{sec:ps1_selection}.


\subsubsection{PSO~J030.1387--17.6238 ($z=6.09$)}
PSO~J030.1387--17.6238 is from our extended selection as it has S/N(\zps)$=8.3$.

\subsubsection{PSO~J173.4601+48.2420  ($z=6.233$)}
This quasar would not have been selected due to its S/N(\yps)$<5$. However, we selected this quasar following the more relaxed selection presented in \cite{banados2015a}, which also required a radio counterpart in the  Faint Images of the Radio Sky at Twenty cm survey (FIRST; \citealt{becker1995}).%

This quasar has a narrow \lya\ line that is not represented in the templates used to estimate the redshift in Section \ref{sec:redshift}. We use the redshift of a narrow absorption \hi\ and \nv\ system at $z=6.233 \pm 0.003$ that might be associated with the host galaxy. The \lya\ line is stronger than all templates except for \textit{strong-\lya} but much narrower than the one in  \textit{strong-\lya}. To estimate the rest-frame magnitudes we fix the redshift to $z=6.233$ and use the \textit{strong-\lya} template.
This radio-loud quasar is further discussed in Section~\ref{sec:radio}.

\section{New radio-loud quasars}\label{sec:radio}

We cross-matched our \nqso{} quasars with the FIRST survey (2014dec17 version), LOFAR Two-metre Sky Survey DR2 \citep[LoTSS-DR2;][]{shimwell2022}, and the first data release of the Rapid ASKAP Continuum Survey \citep[RACS;][]{mcconnell2020,hale2021}  catalogs.
In the FIRST catalog, only P173+48 was detected. This detection was expected as a FIRST counterpart was required for its selection (see Section~\ref{sec:individual_notes}). In the LoTSS-DR2 catalog there are two matches: P173+48 and P207+37 (
these two quasars were also recently reported by \citealt{gloudemans2022}).
In the RACS catalog there are also two matches: P050--18 and P193--02.

We analyzed the FIRST, LoTSS, and RACS images following \cite{banados2015a}. We obtained $1^\prime\times 1^\prime$ ($2^\prime\times 2^\prime$ for RACS)  images and checked whether there were S/N$>3$ detections within $2^{\prime\prime}$ ($4^{\prime\prime}$ for RACS) from the quasars' position.  The FIRST footprint covers 24 of our quasars and our procedure recovered two sources: P173+48 which was already in the FIRST catalog and P193--02 detected at S/N=3.4 with $509\pm151\,\mu$Jy (this last one was also part of the RACS catalog). The LoTSS-DR2 footprint covers 11 of our quasars and we found three detections. In addition to the catalogued P173+48 and P207+37, we recover a S/N$=4.1$ detection of P182+53 with $240\pm60\,\mu$Jy.
The RACS footprint covers 40 of our quasars and our procedure recovered the two catalogued quasars P050--18 and P193--02. We note that in the field of P193--02 there are other three radio sources within 2\arcmin. The two closest ones are clearly associated to a group of foreground galaxies while the one $\sim1\farcm2$ north of the quasar does not have an obvious optical counterpart in \PS.
There is an additional potential  S/N$=3.9$ ($0.94\pm0.24\,$mJy) source 4\farcs1 from P032--17, just outside our required matching radius but still consistent with typical positional uncertainties in the RACS survey. We do not consider P032--17 as radio-loud, however we note that it is an interesting source for future follow-up to confirm whether this radio emission is real and coming from the quasar.

We followed the same procedure to analyse the quick look images of the Very Large Array Sky Survey (VLASS; \citealt{lacy2020}). VLASS provides two-epoch 3\,GHz images of all the sky above declination $-40^{\circ}$, and it therefore covers all of our new quasars. We analyze the two VLASS epochs for each quasar and consider a detection robust only if the measurements are consistent within the two epochs. We recover two VLASS sources: P173+48 and P050--18.  We report the radio properties of these five quasars in Table~\ref{tab:radio-info} and show their radio images in Figure~\ref{fig:radio}.

In Table~\ref{tab:radio-info} we report the radio spectral index ($f_\nu \propto \nu^{\alpha}$) between the 144\,MHz and 1.4\,GHz (\slopelofarfirst), 888\,MHz and 1.4\,GHz (\sloperacsfirst), 888\,MHz and 3\,GHz (\sloperacsvlass), and 1.4\,GHz and 3\,GHz (\slopefirstvlass) when the data exist. We estimate the radio loudness for our sources as $R_{2500}=f_{\nu,5\,\mathrm{GHz}} / f_{\nu,2500\,\mathrm{\AA{}}}$.   We estimate $f_{\nu,2500\,\mathrm{\AA{}}}$ from the rest-frame magnitudes, $m_{2500}$, listed in Table~\ref{tab:qsos-info}.  To estimate the rest-frame flux density at 5\,GHz we extrapolate the observed radio emission using all the spectral slopes listed in Table~\ref{tab:radio-info}, where we report the resulting range of $R_{2500}$. For sources only detected in LoTSS (P182+53 and P207+37) we assume the median spectral index $-0.29$ as measured by \cite{gloudemans2021}. All five sources are classified as radio-loud, i.e., with  $R_{2500}>10$ (see e.g., \citealt{banados2021}), see Table~\ref{tab:radio-info}.

\subsection{Notes on peculiar radio-loud quasars}

\subsubsection{PSO~J173.4601+48.2420 -- blazar candidate}

P173+48 at $z=6.233$ is the brightest radio source in the sample. It is detected in all radio surveys that cover the quasar. Its two VLASS epochs differ by more than 2$\sigma$, which is suggestive of variability. The radio spectral $\slopelofarfirst=-0.16$ and $\slopefirstvlass=-0.41$ (VLASS epoch 1) are remarkably flat (the slightly steeper slope using VLASS epoch 2 $\slopefirstvlass=-0.69$ can be due to variability). Both high variability and a flat radio spectral index ($-0.5<\alpha<0.5$) are indications of a blazar nature, i.e., a quasar with its relativistic jet pointing at a small angle from our line of sight (e.g., \citealt{caccianiga2019,ighina2019}). Monitoring the radio variability, simultaneous constraints on the radio spectral energy distribution, and X-ray observations can help to establish whether this radio-loud quasar is a blazar. If confirmed, this would be the most distant blazar known to date, with the current record at $z=6.1$ \citep{belladitta2020}.

\begin{figure*}
	\centering
	\includegraphics[width=\textwidth]{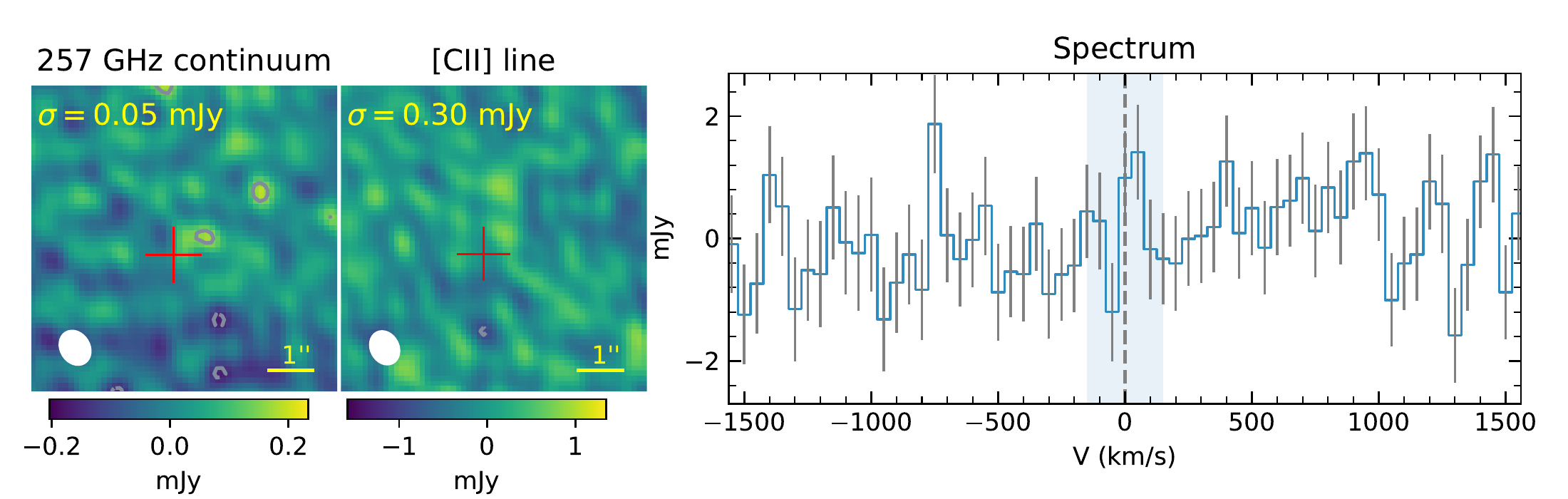}
	\caption{NOEMA observations of the blazar candidate P173+48. \textit{Left}: the 257 GHz continuum image (all datacube excluding the channels around the expected \cii\ emission line (shaded area in the right panel). \textit{Middle}: The integrated image obtained by averaging the channels around the expected \cii\ emission line (shaded area in the right panel). In both left and middle panels, the contour levels are ($-3$,$3$)$\times\sigma$, and the red cross shows the optical position of the quasar. \textit{Right}: Spectrum extracted from the 3.5$\sigma$ peak $\sim1\arcsec$ from the optical position of the quasar (see left panel). The dashed vertical line shows the expected location of the \cii\ line at $z=6.233$.}
	\label{fig:NOEMA_P173+48}
\end{figure*}

We target the \cii\ line (and underlying continuum) of the quasar with the NOEMA interferometer (program W21EC). The observations were carried out on 2021 December 13  with an  on-source time of 7.5~h. We reduced the data with GILDAS\footnote{\url{https://www.iram.fr/IRAMFR/GILDAS}}, following the steps described in \cite{khusanova2022}. We used 3C84 and LKHA101 as bandpass and flux density calibrators, respectively.

Collapsing the entire datacube, except for the line containing channels ($\pm1000$\,km\,s$^{-1}$ from expected \cii\ line location), we find a 3.5$\sigma$  peak $\sim1\arcsec$ offset from the optical position of the quasar (Fig.\ \ref{fig:NOEMA_P173+48}). This signal might be noise given several positive and negative peaks of comparable significance. We extracted a spectrum centered on the 3.5$\sigma$ peak described above and another one centered at the optical position of the quasar, finding no emission line in both cases. We created a line image averaging the channels from $-150$ to $150$\,km\,s$^{-1}$ from the expected \cii\ line. We find no significant detection on this image (middle panel of Fig.~\ref{fig:NOEMA_P173+48}).

We estimate a star-formation rate (SFR) limit based on the continuum flux density, assuming a modified black-body model with $T_{dust}=47$ K and $\beta=1.6$ (standard assumptions, see e.g., \citealt{decarli2018,khusanova2022}).
 Since a tentative detection is present on the continuum image, we use 5$\sigma$ as a conservative upper limit for the SFR, yielding  $<69\,M_{\odot}$\,yr$^{-1}$ (using the conversion of \citealt{kennicutt1998} with a \citealt{Chabrier2003} initial mass function).
 Assuming a \cii\ line width of 300\,km\,s$^{-1}$, the $3\sigma$ upper limit on SFR is 27 $M_{\odot}$\,yr$^{-1}$ (using the conversion of \citealt{delooze2014}).
 Only one other $z\gtrsim 6$ radio-loud quasar has remained undetected in both \cii\ and the underlying continuum with observations of similar depth \citep{khusanova2022}.

\subsubsection{PSO~J193.3992--02.7820 -- radio transient?}

P193--02 has a strong RACS detection, but the relatively weak $3.4\sigma$ FIRST detection implies an ultra-steep radio spectrum with $\alpha\approx -4.6$. Taking the radio spectral index at face value would make it the radio-loudest quasar in the sample with $R_{2500}\approx 1300$. We note, however, that there is a significant difference between the integrated and peak flux densities ($4.05\pm0.07\,$mJy vs.\ $2.86\pm0.02\,$mJy). This discrepancy could mean that the source is extended or that multiple sources contribute
to the integrated flux density. Even if we consider the peak flux density, this results in an ultra-steep spectrum with $\alpha\approx -3.8$ and high radio loudness $R_{2500}\approx 800$. The steep radio spectrum is in agreement with the non-detection in VLASS. Ultra-steep radio sources are expected to be ``lobe dominated'', and it is a signature often used to find high-redshift radio galaxies (e.g., \citealt{saxena2018}). Interestingly, the implied steep radio slope for this source makes it an extreme outlier: there are virtually no quasars or radio galaxies known with such a steep radio spectrum  \citep[c.f.,][]{saxena2018b,sabater2019,banados2021,zajavcek2019}. If such a steep slope continues to lower frequencies without a turnover, we would expect unprecedented radio emission at the Jansky level at around $200\,$MHz. However, the quasar is undetected in the TIFR GMRT Sky Survey \citep[TGSS;][]{intema2017} at 147.5\,MHz. We analyze the TGSS image and obtain a  $3\sigma$ upper limit of 7.1\,mJy (we note that none of the new radio-loud quasars presented in this paper are detected in TGSS). This non-detection implies a substantial turnover between 147.5 and 888\,MHz.

Another possibility is that the strong RACS detection of P193--02 was a radio transient \cite[e.g.,][]{mooley2016}. A flux density of 4.1\,mJy translates to a specific luminosity of $L_\nu=2.2\times 10^{33}\lnu$ at the redshift of the quasar, which is brighter than the most radio luminous supernovae or gamma-ray bursts ever observed but consistent with an AGN flare \citep[][]{pietka2015,nyland2020}. We note that the strong RACS emission detected on 2020 May 1 happened  between the non-detections in the two VLASS epochs (2019 April 21 and 2021 Dec 12). The whole period is just about five months in the quasar's rest frame. Assuming a more realistic steep radio spectral index $\alpha=-1$, given the RACS detection we would expect a $1.2\,$mJy source in VLASS. The fact that we do not detect it implies that the source should have increased its luminosity by at least a factor of three in $\sim$2-month rest-frame and faded by the same factor in $\sim$3-month rest-frame. Such a rapid variability suggests blazar-like activity \citep{pietka2015}.
Assuming that the FIRST flux density is representative of its steady state and assuming $\alpha=-0.63$ (median value used in \cite{banados2021} for objects without a measured radio slope), P193--02 would still be classified as radio-loud with $R_{2500}\approx 100$. Another epoch at 888\,MHz and/or higher-resolution/deeper observations are required to understand this source better.


\begin{figure*}[h!]
\centering
\includegraphics[scale=0.8]{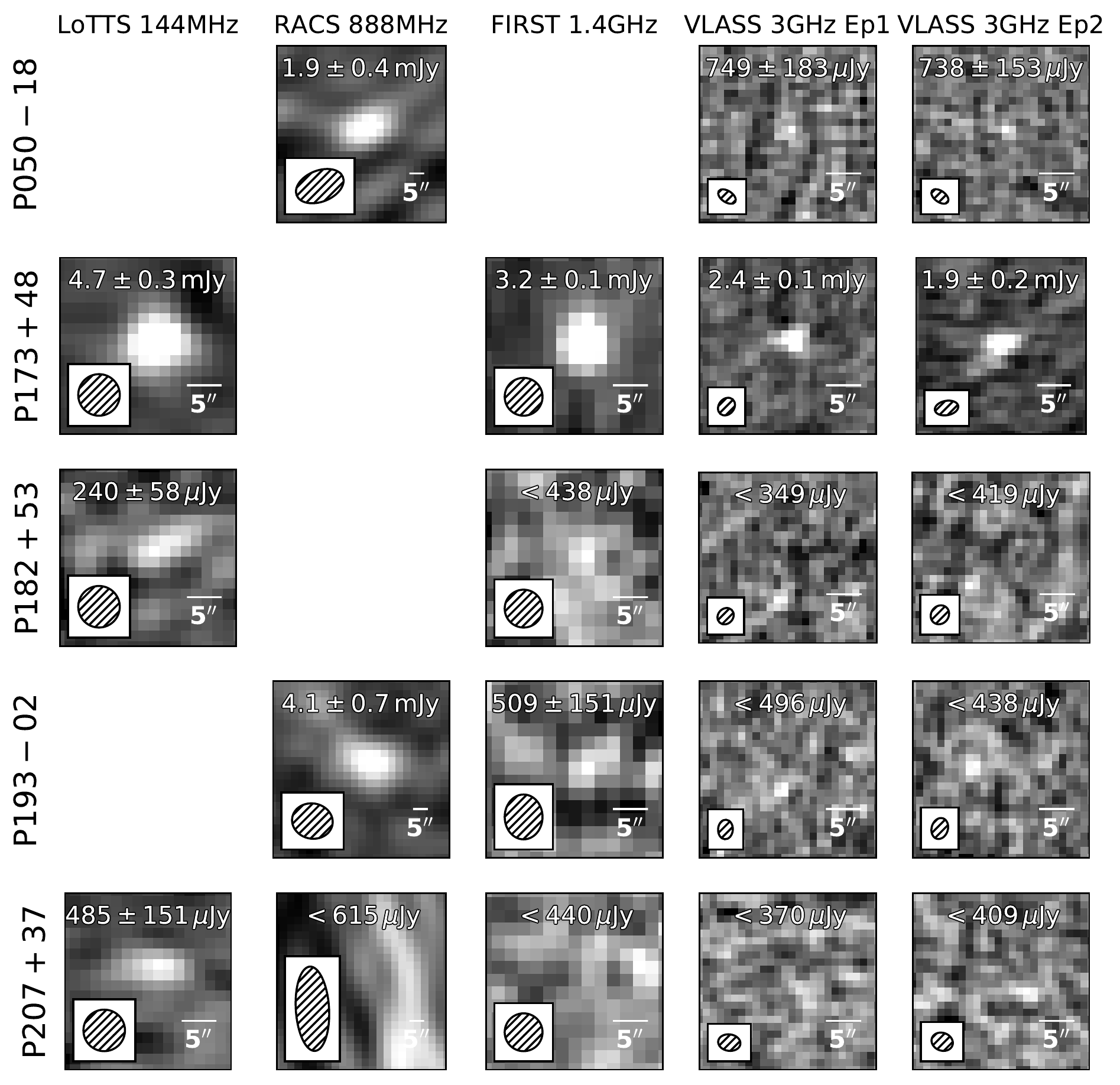}
 \caption{$25^{\prime\prime}\times25^{\prime\prime}$ ($60^{\prime\prime}\times60^{\prime\prime}$ for RACS) radio images for the  quasars detected in at least one of the following radio surveys: LoTSS-DR2 (left column), RACS (second column), FIRST (third column) or both VLASS epochs (fourth and fifth columns). The measured flux density (or limits) are shown at the top of each postage stamp, and the beam sizes are indicated at the bottom left of each image. See section \ref{sec:radio}.
 \label{fig:radio}}
 \end{figure*}

\section{Summary}\label{sec:summary}

In this paper, we present the discovery of \nqso{} quasars at $5.6<z<6.5$, continuing the work from \cite{banados2016} and \cite{venemans2015b} to select high-redshift quasars with the \PS\ and VIKING surveys, respectively. Our selection function for $z\sim6$ \PS\ quasars is discussed and is used to derive the $z\sim 6$ quasar luminosity function in the companion paper by \cite{schindler2022}.

The quasars presented here show a range of properties, including objects with weak emission line quasars and BAL quasars, which seem  more common at $z\sim6$ than at lower redshifts (see e.g., \citealt{banados2016,bischetti2022}). 10\% of the quasars presented here are radio-loud, including one blazar candidate. This percentage is in line with the expectations \citep[e.g.,][]{banados2015a,liu2021}. We also present a $z=6.41$ quasar that could be the second gravitationally lensed quasar known at $z>5$. Multi-wavelength follow-up observations of the quasars presented in this sample are required to understand
their nature and physical properties better.

 Two of the VIKING quasars could have been selected by the \PS\ selection, if we had relaxed our S/N requirements. The additional discoveries just below our nominal S/N cuts demonstrate that there are still more high-redshift quasars to be found in the Pan-STARRS1 survey (Section~\ref{sec:notes_extended}), likely at the expense of significantly larger contamination. Given the existence of large sky radio surveys, including recent ones such as LoTSS, RACS, and VLASS (see Fig.~\ref{fig:radio}), relaxing our S/N and/or color requirements described in Section~\ref{sec:selection} but requiring a radio detection could be a promising way to expose some of the quasar populations we are currently missing. Eventually, the advent of large multi-object spectroscopic surveys such as 4MOST \citep{dejong2019} and DESI \citep{chaussidon2022} will be fundamental to maximize the discovery of quasars that are located in color space dominated by other astronomical sources and/or that are at the limits of current selections.


\vspace{4mm}
{\small
EB would like to thank to all the telescope operators and observatory staff who make the many nights at the telescopes discovering these quasars enjoyable and successful.  We thank Linhua Jiang for providing the discovery spectrum of P218+04.
EB also thanks Alex Ji, Silvia Belladitta, Dillon Dong, Andrew Newman,
Konstantina Boutsia, Michael Rauch, Peter Boorman, Marianne Heida, George Lansbury, Adric Riedel, Emmanuel Momjian, Fred Davies for insightful discussions and/or support in some of the observing runs.
AP acknowledges support from Fondazione Cariplo grant no. 2020-0902.
RAM acknowledges support from the ERC Advanced Grant 740246 (Cosmic\_Gas).
EPF is supported by the international Gemini Observatory, a program of NSF’s NOIRLab, which is managed by the Association of Universities for Research in Astronomy (AURA) under a cooperative agreement with the National Science Foundation, on behalf of the Gemini partnership of Argentina, Brazil, Canada, Chile, the Republic of Korea, and the United States of America.
GN acknowledges funding support from the Natural Sciences and Engineering Research Council (NSERC) of Canada through a Discovery Grant and Discovery Accelerator Supplement, and from the Canadian Space Agency through grant 18JWST-GTO1.
JS acknowledges support from the JPL RTD program.

Part of this work has been made possible thanks to Netherlands Research School for Astronomy instrumentation grant for the AstroWISE information system.

The LBT is an international collaboration among institutions in the United States, Italy and Germany. The LBT Corporation partners are: The University of Arizona on behalf of the Arizona university system; Istituto Nazionale di Astrofisica, Italy;  LBT Beteiligungsgesellschaft, Germany, representing the Max Planck Society, the Astrophysical Institute Potsdam, and Heidelberg University; The Ohio State University; The Research Corporation, on behalf of The University of Notre Dame, University of Minnesota and University of Virginia.
%
This paper used data obtained with the MODS spectrograph built with
funding from NSF grant AST-9987045 and the NSF Telescope System
Instrumentation Program (TSIP), with additional funds from the Ohio
Board of Regents and the Ohio State University Office of Research.
%
This paper includes data gathered with the 6.5 meter Magellan Telescopes located at Las Campanas Observatory, Chile.

This work is based on observations made with ESO Telescopes at the La Silla Paranal Observatory under programs
106.20WQ,
092.A-0339(A),
096.A-0420(A),
096.A-0420(A),
092.A-0339(A),
094.A-0053(A),
094.A-0053(A),
093.A-0863(A),
%
%
091.A-0290(A),
092.A-0150(B),
093.A-0574(A),
095.A-0535(B),
097.A-0094(B),
0100.A-0215(B),
0104.A-0662(A),
099.A-0424(B),
0101.A-0135(A),
106.20WJ.001,
097.A-9001(A).
%

This work was enabled by observations made from the Gemini North and Keck elescopes, located within the Maunakea Science Reserve and adjacent to the summit of Maunakea. We are grateful for the privilege of observing the Universe from a place that is unique in both its astronomical quality and its cultural significance.

%
%

Based on observations obtained at the international Gemini Observatory (under program: GN-2022A-Q-411), a program of NSF’s NOIRLab, which is managed by the Association of Universities for Research in Astronomy (AURA) under a cooperative agreement with the National Science Foundation on behalf of the Gemini Observatory partnership: the National Science Foundation (United States), National Research Council (Canada), Agencia Nacional de Investigaci\'{o}n y Desarrollo (Chile), Ministerio de Ciencia, Tecnolog\'{i}a e Innovaci\'{o}n (Argentina), Minist\'{e}rio da Ci\^{e}ncia, Tecnologia, Inova\c{c}\~{o}es e Comunica\c{c}\~{o}es (Brazil), and Korea Astronomy and Space Science Institute (Republic of Korea).

Based on observations carried out with the IRAM Interferometer NOEMA. IRAM is supported by INSU/CNRS (France), MPG (Germany) and IGN (Spain).
%
%

The Pan-STARRS1 Surveys (PS1) and the PS1 public science archive have been made possible through contributions by the Institute for Astronomy, the University of Hawaii, the Pan-STARRS Project Office, the Max-Planck Society and its participating institutes, the Max Planck Institute for Astronomy, Heidelberg and the Max Planck Institute for Extraterrestrial Physics, Garching, The Johns Hopkins University, Durham University, the University of Edinburgh, the Queen's University Belfast, the Harvard-Smithsonian Center for Astrophysics, the Las Cumbres Observatory Global Telescope Network Incorporated, the National Central University of Taiwan, the Space Telescope Science Institute, the National Aeronautics and Space Administration under Grant No. NNX08AR22G issued through the Planetary Science Division of the NASA Science Mission Directorate, the National Science Foundation Grant No. AST-1238877, the University of Maryland, Eotvos Lorand University (ELTE), the Los Alamos National Laboratory, and the Gordon and Betty Moore Foundation.

LOFAR data products were provided by the LOFAR Surveys Key Science project (LSKSP; \url{https://lofar-surveys.org/}) and were derived from observations with the International LOFAR Telescope (ILT). LOFAR (van Haarlem et al. 2013) is the Low Frequency Array designed and constructed by ASTRON. It has observing, data processing, and data storage facilities in several countries, which are owned by various parties (each with their own funding sources), and which are collectively operated by the ILT foundation under a joint scientific policy. The efforts of the LSKSP have benefited from funding from the European Research Council, NOVA, NWO, CNRS-INSU, the SURF Co-operative, the UK Science and Technology Funding Council and the Jülich Supercomputing Centre.

The ASKAP radio telescope is part of the Australia Telescope National Facility which is managed by Australia’s national science agency, CSIRO. Operation of ASKAP is funded by the Australian Government with support from the National Collaborative Research Infrastructure Strategy. ASKAP uses the resources of the Pawsey Supercomputing Research Centre. Establishment of ASKAP, the Murchison Radio-astronomy Observatory and the Pawsey Supercomputing Research Centre are initiatives of the Australian Government, with support from the Government of Western Australia and the Science and Industry Endowment Fund. We acknowledge the Wajarri Yamatji people as the traditional owners of the Observatory site. This paper includes archived data obtained through the CSIRO ASKAP Science Data Archive, CASDA (\url{https://data.csiro.au}).
}

%


\software{astropy \citep{2013A&A...558A..33A,2018AJ....156..123A},
          matplotlib \citep{hunter2007},
          numpy \citep{harris2020},
          linetools (\url{https://github.com/linetools}),
          SciPy \citep{virtanen2020},
TOPCAT \citep[][\url{http://www.starlink.ac.uk/topcat/}]{taylor2005},
GILDAS (\url{https://www.iram.fr/IRAMFR/GILDAS})
}





\bibliographystyle{aasjournal}



\clearpage
\input{tab1_arxiv}
\startlongtable
\input{tab2}
\startlongtable
\begin{longrotatetable}
\input{tab3}
\end{longrotatetable}
\input{tab4}
\end{CJK*}
\end{document}

%% file: tab1_arxiv.tex
\startlongtable
\begin{deluxetable*}{lcrrrrr}
\tablecolumns{5}
\tablewidth{0pc}
\tablecaption{Discovery spectroscopic observations of the new quasars presented in this work, sorted by right ascension. The PS1 quasars are shown first and the VIKING quasars are at the end of the table. \label{tab:spectroscopy}}
 \tablehead{
\colhead{Quasar}    &  \colhead{Date} &   \colhead{Telescope/Instrument}   & \colhead{Exposure Time} &
\colhead{Slit Width} & \colhead{QLF?}
}
\startdata
PSO~J000.0416--04.2739   & 2021 Nov 30        & P200/DBSP       & 1200\,s       & 1\farcs5 & T \\
PSO~J000.0805-18.2360 & 2017 May 15          &  Magellan/LDSS3      & 1800\,s       & 1\farcs0 & F \\
                 & 2018 Jun 4          &  Magellan/LDSS3      & 1500\,s       & 1\farcs0 &  \\
                 & 2022 Sep 17          &  P200/DBSP      & 1200\,s       & 1\farcs5 &  \\
PSO~J001.6889--28.6783 & 2017 Aug 05          &  Magellan/LDSS3      & 1800\,s       & 1\farcs0 & F \\
PSO~J002.5429+03.0632   & 2021 Dec 5        & P200/DBSP       & 1300\,s       & 1\farcs5 & T \\
PSO~J008.6083+10.8518  & 2016 Nov 30  	  & LBT/MODS        & 1800\,s	 & 1\farcs2 & F	\\
PSO~J017.0691--11.9919  & 2016 Nov 26  	  & LBT/MODS        & 1800\,s	 & 1\farcs2 & T	\\
PSO~J030.1387--17.6238 & 2021 Nov 17--18         &  NTT/EFOSC      & 9000\,s       & 1\farcs0--1\farcs5 & F \\
                 & 2022 Sep 17          &  P200/DBSP      & 1200\,s      & 1\farcs5 &  \\
PSO~J032.91882--17.0746 & 2017 Sep 26          &  Magellan/LDSS3      & 700\,s       & 1\farcs0 & F \\
PSO~J038.1914--18.5735 & 2017 Sep 26          &  Magellan/LDSS3      & 1800\,s       & 1\farcs0 & T \\
PSO~J043.1111--02.6224 & 2021 Nov 19         &  NTT/EFOSC      & 2580\,s       & 1\farcs5 & F \\
                 & 2022 Sep 17          &  P200/DBSP     & 1200\,s       & 1\farcs5 &  \\
PSO~J050.5605--18.6881 & 2017 Sep 26          &  Magellan/LDSS3      & 1800\,s       & 1\farcs0 & F \\
PSO~J065.9589+01.7235 & 2017 Mar 06          &  Magellan/LDSS3      & 900\,s       & 1\farcs0 & F \\  but in PV3 zy<0.5
PSO~J072.5825--07.8918   & 2021 Nov 30        & P200/DBSP       & 2400\,s       & 1\farcs5 & T \\
PSO~J076.2344--10.8878   & 2022 Sep 28        & P200/DBSP       & 1020\,s       & 1\farcs5 & T \\
PSO~J119.0932+02.3056 & 2017 Mar 06          &  Magellan/LDSS3      & 600\,s       & 1\farcs0 & F \\
                      & 2018 Jan 25          &  Magellan/LDSS3      & 900\,s       & 1\farcs0 &  \\
PSO~J124.0032+12.9989   & 2021 Dec 5        & P200/DBSP       & 1200\,s       & 1\farcs5 & T \\
PSO~J127.0558+26.5654  & 2021 Nov 10  	  & LBT/MODS        &1800\,s	 & 1\farcs2 & T	\\
PSO~J142.3990--11.3604 & 2017 Mar 06          &  Magellan/LDSS3      & 900\,s       & 1\farcs0 & F \\
PSO~J148.4829+69.1812   & 2021 Dec 5        & P200/DBSP       & 900\,s       & 1\farcs5& T \\
PSO~J151.8186-16.1855 & 2017 May 15          &  Magellan/LDSS3      & 1200\,s       & 1\farcs0 & F \\
                 & 2018 Jun 4          &  Magellan/LDSS3      & 1200\,s       & 1\farcs0 &  \\
PSO~J156.4466+38.9573   & 2021 Dec 5        & P200/DBSP       & 400\,s       & 1\farcs5 & T \\
PSO~J158.6937--14.4210			  & 2016 Nov 27        & Keck/LRIS       & 900\,s          & 1\farcs0 & F\\
PSO~J169.1406+58.8894  & 2017 Apr 19  	  & LBT/MODS        & 840\,s	 & 1\farcs2 & T	\\
PSO~J170.8326+20.2082 & 2022 June 13          &  Gemini-N/GMOS      & 1200\,s       & 1\farcs5 & F  \\
PSO~J173.3198--06.9458 & 2017 Mar 06          &  Magellan/LDSS3      & 600\,s       & 1\farcs0 & F \\
PSO~J173.4601+48.2420  & 2021 May 12  	  & LBT/MODS        & 4800\,s	 & 1\farcs2 & F	\\
PSO~J175.4294+71.3236  & 2021 Dec 5        & P200/DBSP       & 600\,s       & 1\farcs5 & T \\
PSO~J178.3733+28.5075   & 2021 Nov 30        & P200/DBSP       & 600\,s       & 1\farcs5 & T \\
PSO~J182.3121+53.4633  & 2021 Dec 6        & P200/DBSP       & 1500\,s       & 1\farcs5 & T \\
PSO~J193.3992--02.7820 & 2018 Jan 25          &  Magellan/LDSS3      & 900\,s       & 1\farcs0 & T \\
PSO~J196.3476+15.3899  & 2021 Dec 5        & P200/DBSP       & 1200\,s       & 1\farcs5 & T \\
PSO~J197.8675+45.8040  & 2018 Jun 10        & P200/DBSP       & 1800\,s       & 1\farcs5 & T \\
PSO~J207.5983+37.8099  & 2017 Apr 19  	  & LBT/MODS        & 840\,s	 & 1\farcs2 & T	\\
                       & 2017 Apr 28  	  & Keck/LRIS        & 1200\,s	 & 1\farcs0 & 	\\
PSO~J207.7780--21.1889 & 2017 Mar 06          &  Magellan/LDSS3      & 700\,s       & 1\farcs0 & F \\
                       & 2019 Jun 09          &  LBT/MODS      & 1800\,s       & 1\farcs2 &  \\
PSO~J209.3825--08.7171 & 2017 Aug 08          &  Magellan/LDSS3      & 1200\,s       & 1\farcs0 & T \\
PSO~J215.4303+26.5325  & 2016 May 05	      & MMT/Red Channel & 2400\,s       & 1\farcs0    & F    \\
PSO~J218.3967+28.3306   & 2021 May 13  	  & LBT/MODS        & 1800\,s	 & 1\farcs2 & T	\\ %
PSO~J218.7714+04.8189   & 2021 Jul 13     & P200/DBSP       & 3600\,s       & 1\farcs5 & T \\
PSO~J224.6506+10.2137				  & 2022 Mar 6        & Keck/LRIS       & 300\,s          & 1\farcs0 & T\\
PSO~J228.7029+01.3811 & 2021 Feb 19         & VLT/FORS2       & 450\,s       & 1\farcs3 & F\\
PSO~J261.1247+37.3060			  & 2022 Mar 6        & Keck/LRIS       & 300\,s          & 1\farcs0 & T\\
PSO~J265.9298+41.4139  & 2017 Apr 19  	  & LBT/MODS        & 840\,s	 & 1\farcs2 & T	\\
PSO~J271.4455+49.3067  & 2017 Apr 19  	  & LBT/MODS        & 840\,s	 & 1\farcs2 & T	\\ %
PSO~J281.3361+53.7631  & 2020 Oct 22  	  & LBT/MODS        & 1200\,s	 & 1\farcs2 & T	\\ %
PSO~J288.6476+63.2479    & 2022 Sep 17     & P200/DBSP      & 2400\,s    & 1\farcs5 & T \\
PSO~J306.3512--04.8227   & 2022 Sep 17     & P200/DBSP      & 1320\,s    & 1\farcs5 & T \\
PSO~J307.7635--05.1958   & 2022 Sep 17     & P200/DBSP      & 1290\,s    & 1\farcs5 & T \\
PSO~J334.0181--05.0048 & 2017 Sep 26          &  Magellan/LDSS3      & 1200\,s       & 1\farcs0 & T \\
\hline
VIK~J0046--2837 & 2014 Oct 28        & VLT/FORS2       & 1362\,s      & 1\farcs3  & F\\
VIK~J0224--3435 & 2014 Nov 14        & VLT/FORS2       & 1482\,s     & 1\farcs3  & F\\
VIK~J1152+0055  & 2014 Apr 30        & VLT/FORS2       & 2615\,s      & 1\farcs3  & F\\
VIK~J2211--3206 & 2013 Dec 19       & VLT/FORS2       & 1482\,s     & 1\farcs3  & F\\
VIK~J2227--3323 & 2015 Nov 05        & VLT/FORS2       & 2600\,s      & 1\farcs3  & F\\
VIK~J2315--2856 & 2015 Dec 10       & VLT/FORS2       & 1500\,s     & 1\farcs3  & F\\
VIK~J2318--3029 & 2013 Dec 21        & VLT/FORS2       & 1482\,s      & 1\farcs3  & F\\
\enddata
\tablecomments{For full coordinates and redshifts see Table \ref{tab:qsos-info}. The last column indicates whether the quasars are part of the $z\sim 6$ quasar luminosity function presented in \cite{schindler2022}.
}
\end{deluxetable*}

%% file: tab2.tex
\begin{deluxetable*}{ccccccccccc}

    \centerwidetable
\movetabledown=5mm
    \tablewidth{0pt}
    \tabletypesize{\footnotesize}
    
\tablecaption{Properties of the quasars discovered in this work, sorted by right ascension. \label{tab:qsos-info}}
\tablehead{\colhead{Quasar} & \colhead{R.A.} & \colhead{Decl.} & \colhead{$z$} & \colhead{$z$ method} & \colhead{$z$ ref.} & \colhead{$m_{1450}$} & \colhead{$M_{1450}$} & \colhead{$m_{2500}$} & \colhead{Best} & \colhead{Notes}\\ \colhead{ } & \colhead{ICRS} & \colhead{ICRS} & \colhead{ } & \colhead{ } & \colhead{ } & \colhead{$\mathrm{mag}$} & \colhead{$\mathrm{mag}$} & \colhead{$\mathrm{mag}$} & \colhead{template} & \colhead{ }}
\startdata
P000--04 & 00:00:09.99 & $-$04:16:26.05 & 5.77 & template & 1 & 20.79 & $-25.84$ & 21.33 & \textit{weak-\lya} &  \\
P000--18 & 00:00:19.33 & $-$18:14:09.64 & 5.9 & template & 1 & 21.48 & $-25.19$ & 22.02 & \textit{strong-\lya} &  \\
P001--28 & 00:06:45.36 & $-$28:40:42.05 & 5.95 & template & 1 & 21.07 & $-25.61$ & 21.75 & \textit{selsing2016} &  \\
P002+03 & 00:10:10.30 & $+$03:03:47.57 & 5.64 & template & 1 & 20.92 & $-25.67$ & 21.46 & \textit{weak-\lya} & low S/N-BAL \\
P008+10 & 00:34:26.00 & $+$10:51:06.70 & 5.98 & template & 1 & 20.62 & $-26.07$ & 21.16 & \textit{yang2021} &  \\
J0046--2837 & 00:46:23.65 & $-$28:37:47.68 & 6.02 & template & 1 & 21.34 & $-25.36$ & 21.88 & \textit{yang2021} &  \\
P017--11 & 01:08:16.60 & $-$11:59:31.00 & 5.8 & template & 1 & 20.70 & $-25.94$ & 21.31 & \textit{strong-\lya} &  \\
P030--17 & 02:00:33.30 & $-$17:37:26.00 & 6.09 & template & 1 & 21.36 & $-25.36$ & 21.90 & \textit{strong-\lya} &  \\
P032--17 & 02:11:40.52 & $-$17:04:28.74 & 5.99 & template & 1 & 20.89 & $-25.80$ & 21.43 & \textit{weak-\lya} &  \\
J0224--3435 & 02:24:56.75 & $-$34:35:23.00 & 5.78 & template & 1 & 21.77 & $-24.86$ & 22.31 & \textit{weak-\lya} &  \\
P038--18 & 02:32:45.94 & $-$18:34:24.61 & 5.68 & template & 1 & 20.57 & $-26.04$ & 21.11 & \textit{weak-\lya} &  \\
P043--02 & 02:52:26.68 & $-$02:37:20.70 & 6.17 & template & 1 & 21.05 & $-25.69$ & 21.72 & \textit{vandenberk2001} &  \\
P050--18 & 03:22:14.54 & $-$18:41:17.43 & 6.09 & template & 1 & 20.45 & $-26.27$ & 20.99 & \textit{weak-\lya} & radio-loud \\
P065+01 & 04:23:50.15 & $+$01:43:24.73 & 5.79 & template & 1 & 20.08 & $-26.56$ & 20.62 & \textit{weak-\lya} & BAL \\
P072--07 & 04:50:19.81 & $-$07:53:30.59 & 5.75 & template & 1 & 20.97 & $-25.66$ & 21.51 & \textit{strong-\lya} &  \\
P076--10 & 05:04:56.27 & $-$10:53:16.09 & 5.93 & template & 1 & 20.55 & $-26.13$ & 21.09 & \textit{weak-\lya} &  \\
P119+02 & 07:56:22.38 & $+$02:18:20.16 & 5.73 & template & 1 & 20.42 & $-26.20$ & 20.96 & \textit{weak-\lya} &  \\
P124+12 & 08:16:00.79 & $+$12:59:56.31 & 5.8 & template & 1 & 21.05 & $-25.59$ & 21.59 & \textit{median-\lya} &  \\
P127+26 & 08:28:13.41 & $+$26:33:55.49 & 6.14 & template & 1 & 20.52 & $-26.21$ & 21.06 & \textit{median-\lya} &  \\
P142--11 & 09:29:35.77 & $-$11:21:37.59 & 5.61 & template & 1 & 20.60 & $-25.99$ & 21.14 & \textit{weak-\lya} &  \\
P148+69 & 09:53:55.90 & $+$69:10:52.62 & 5.84 & template & 1 & 20.19 & $-26.46$ & 20.73 & \textit{weak-\lya} &  \\
P151--16 & 10:07:16.49 & $-$16:11:07.94 & 5.84 & template & 1 & 21.45 & $-25.20$ & 22.12 & \textit{vandenberk2001} &  \\
P156+38 & 10:25:47.19 & $+$38:57:26.30 & 5.75 & template & 1 & 19.85 & $-26.78$ & 20.39 & \textit{strong-\lya} &  \\
P158--14 & 10:34:46.51 & $-$14:25:15.85 & 6.0685 & \cii & 3 & 19.39 & $-27.32$ & 19.93 & \textit{median-\lya} &  \\
P169+58 & 11:16:33.76 & $+$58:53:22.19 & 5.73 & template & 1 & 20.59 & $-26.03$ & 21.13 & \textit{median-\lya} &  \\
P170+20 & 11:23:19.84 & $+$20:12:29.79 & 6.41 & template & 1 & 20.70 & $-26.10$ & 21.24 & \textit{yang2021} & lensed? \\
P173--06 & 11:33:16.78 & $-$06:56:44.85 & 5.77 & template & 1 & 20.66 & $-25.97$ & 21.20 & \textit{strong-\lya} & BAL \\
P173+48 & 11:33:50.44 & $+$48:14:31.21 & 6.233 & absorption & 1 & 21.63 & $-25.12$ & 22.17 & \textit{strong-\lya} & radio-loud \\
P175+71 & 11:41:43.07 & $+$71:19:25.03 & 5.86 & template & 1 & 20.71 & $-25.95$ & 21.25 & \textit{yang2021} &  \\
J1152+0055 & 11:52:21.27 & $+$00:55:36.69 & 6.3643 & \cii & 2 & 21.74 & $-25.05$ & 22.41 & \textit{vandenberk2001} &  \\
P178+28 & 11:53:29.60 & $+$28:30:27.11 & 5.68 & template & 1 & 19.84 & $-26.77$ & 20.38 & \textit{weak-\lya} &  \\
P182+53 & 12:09:14.93 & $+$53:27:48.05 & 5.99 & template & 1 & 21.05 & $-25.64$ & 21.58 & \textit{yang2021} & radio-loud \\
P193--02 & 12:53:35.82 & $-$02:46:55.29 & 5.8 & template & 1 & 21.21 & $-25.43$ & 21.75 & \textit{median-\lya} & radio-loud \\
P196+15 & 13:05:23.43 & $+$15:23:23.68 & 5.69 & template & 1 & 20.44 & $-26.17$ & 20.98 & \textit{weak-\lya} & low S/N \\
P197+45 & 13:11:28.20 & $+$45:48:14.69 & 5.66 & template & 1 & 20.87 & $-25.73$ & 21.41 & \textit{weak-\lya} & low S/N \\
P207+37 & 13:50:23.60 & $+$37:48:35.67 & 5.69 & template & 1 & 20.76 & $-25.85$ & 21.29 & \textit{weak-\lya} & radio-loud \\
P207--21 & 13:51:06.72 & $-$21:11:20.27 & 5.81 & template & 1 & 20.69 & $-25.95$ & 21.23 & \textit{strong-\lya} & BAL \\
P209--08 & 13:57:31.82 & $-$08:43:01.74 & 5.77 & template & 1 & 21.44 & $-25.19$ & 21.98 & \textit{strong-\lya} &  \\
P215+26 & 14:21:43.29 & $+$26:31:57.14 & 6.28 & template & 1 & 20.40 & $-26.37$ & 20.94 & \textit{strong-\lya} & BAL \\
P218+28 & 14:33:35.22 & $+$28:19:50.60 & 5.91 & template & 1 & 20.82 & $-25.85$ & 21.36 & \textit{median-\lya} &  \\
P218+04 & 14:35:05.15 & $+$04:49:08.32 & 6.14 & template & 1 & 20.83 & $-25.90$ & 21.50 & \textit{vandenberk2001} & BAL \\
P224+10 & 14:58:36.16 & $+$10:12:49.67 & 5.6 & template & 1 & 19.68 & $-26.90$ & 20.22 & \textit{weak-\lya} & BAL \\
P228+01 & 15:14:48.70 & $+$01:22:52.25 & 5.76 & template & 1 & 21.38 & $-25.25$ & 22.05 & \textit{selsing2016} &  \\
P261+37 & 17:24:29.94 & $+$37:18:21.80 & 5.76 & template & 1 & 21.03 & $-25.60$ & 21.70 & \textit{vandenberk2001} &  \\
P265+41 & 17:43:43.15 & $+$41:24:50.22 & 6.0263 & \cii & 3 & 20.76 & $-25.94$ & 21.29 & \textit{yang2021} & BAL \\
P271+49 & 18:05:46.93 & $+$49:18:24.23 & 5.74 & template & 1 & 20.38 & $-26.24$ & 21.05 & \textit{vandenberk2001} &  \\
P281+53 & 18:45:20.68 & $+$53:45:47.34 & 6.18 & template & 1 & 20.09 & $-26.65$ & 20.63 & \textit{strong-\lya} & BAL \\
P288+63 & 19:14:35.44 & $+$63:14:52.54 & 5.96 & template & 1 & 20.64 & $-26.04$ & 21.18 & \textit{yang2021} &  \\
P306-04 & 20:25:24.31 & $-$04:49:21.88 & 5.84 & template & 1 & 20.71 & $-25.94$ & 21.25 & \textit{yang2021} &  \\
P307-05 & 20:31:03.26 & $-$05:11:45.20 & 5.8 & template & 1 & 20.86 & $-25.78$ & 21.53 & \textit{vandenberk2001} &  \\
J2211--3206 & 22:11:12.39 & $-$32:06:12.95 & 6.3394 & \cii & 2 & 18.78 & $-28.00$ & 19.45 & \textit{strong-\lya} & BAL \\
P334--05 & 22:16:04.36 & $-$05:00:17.58 & 6.15 & template & 1 & 21.07 & $-25.66$ & 21.61 & \textit{median-\lya} &  \\
J2227--3323 & 22:27:18.58 & $-$33:23:35.02 & 6.12 & template & 1 & 21.32 & $-25.41$ & 21.86 & \textit{weak-\lya} &  \\
J2315--2856 & 23:15:07.39 & $-$28:56:17.40 & 5.89 & template & 1 & 21.31 & $-25.35$ & 21.85 & \textit{strong-\lya} &  \\
J2318--3029 & 23:18:33.10 & $-$30:29:33.36 & 6.1456 & \cii & 4 & 20.34 & $-26.39$ & 20.88 & \textit{median-\lya} &
\enddata
\tablerefs{(1) This work, (2) \cite{decarli2018}, (3) \cite{eilers2020}, (4) \cite{venemans2020}}\tablecomments{
The typical redshift uncertainties for estimates based on ``$z$ method $=$ template'' is $<0.03$ (see Section \ref{sec:redshift})
}

\end{deluxetable*}

%% file: tab3.tex
\begin{deluxetable*}{cccccccccccccccccc}

    \centerwidetable
\movetabledown=5mm
    \tablewidth{0pt}
    \tabletypesize{\tiny}
    
\tablecaption{Photometry of the quasars discovered in this work, sorted by right ascension. \label{tab:qsos-phot}}
\tablehead{\colhead{Quasar} & \colhead{\gps} & \colhead{\rps} & \colhead{\ips} & \colhead{\zps} & \colhead{\yps} & \colhead{$I$} & \colhead{$I_{\mathrm{ref}}$} & \colhead{$Z$} & \colhead{$Z_{\mathrm{ref}}$} & \colhead{$Y$} & \colhead{$Y_{\mathrm{ref}}$} & \colhead{$J$} & \colhead{$J_{\mathrm{ref}}$} & \colhead{$H$} & \colhead{$H_{\mathrm{ref}}$} & \colhead{$K$} & \colhead{$K_{\mathrm{ref}}$}\\ \colhead{ } & \colhead{$\mathrm{mag}$} & \colhead{$\mathrm{mag}$} & \colhead{$\mathrm{mag}$} & \colhead{$\mathrm{mag}$} & \colhead{$\mathrm{mag}$} & \colhead{$\mathrm{mag}$} & \colhead{ } & \colhead{$\mathrm{mag}$} & \colhead{ } & \colhead{$\mathrm{mag}$} & \colhead{ } & \colhead{$\mathrm{mag}$} & \colhead{ } & \colhead{$\mathrm{mag}$} & \colhead{ } & \colhead{$\mathrm{mag}$} & \colhead{ }}
\startdata
P000--04 & $>23.84$ & $>23.44$ & $>23.92$ & $20.70 \pm 0.06$ & $20.50 \pm 0.11$ & -- & -- & $20.74 \pm 0.01$ & 18 & $20.74 \pm 0.25$ & 22 & $20.35 \pm 0.19$ & 22 & -- & -- & -- & -- \\
P000--18 & $>23.82$ & $>23.75$ & $23.54 \pm 0.34$ & $21.09 \pm 0.07$ & $21.02 \pm 0.17$ & $23.24 \pm 0.18$ & 8 & -- & -- & $21.36 \pm 0.26$ & 22 & $21.32 \pm 0.15$ & 11 & -- & -- & -- & -- \\
P001--28 & $>23.49$ & $>23.67$ & $>23.42$ & $21.03 \pm 0.07$ & $21.17 \pm 0.19$ & -- & -- & -- & -- & -- & -- & $21.47 \pm 0.11$ & 15 & $21.31 \pm 0.15$ & 15 & -- & -- \\
P002+03 & $>23.79$ & $>23.78$ & $23.05 \pm 0.22$ & $20.83 \pm 0.07$ & $20.89 \pm 0.16$ & $21.90 \pm 0.04$ & 17 & $21.08 \pm 0.02$ & 18 & $20.84 \pm 0.18$ & 21 & $19.97 \pm 0.11$ & 21 & $19.30 \pm 0.08$ & 21 & $19.21 \pm 0.07$ & 21 \\
P008+10 & $>23.56$ & $>23.53$ & $22.80 \pm 0.25$ & $20.74 \pm 0.07$ & $20.38 \pm 0.10$ & -- & -- & $20.34 \pm 0.02$ & 18 & $20.28 \pm 0.05$ & 16 & $20.22 \pm 0.14$ & 21 & $20.12 \pm 0.19$ & 21 & $19.94 \pm 0.18$ & 21 \\
J0046--2837 & $>23.88$ & $>23.44$ & $>23.70$ & $21.47 \pm 0.12$ & $21.01 \pm 0.19$ & $23.65 \pm 0.06$ & 3 & $21.44 \pm 0.06$ & 23 & $21.25 \pm 0.11$ & 23 & $20.92 \pm 0.08$ & 23 & $20.66 \pm 0.09$ & 23 & $20.24 \pm 0.09$ & 23 \\
P017--11 & $>23.81$ & $>23.55$ & $23.19 \pm 0.27$ & $20.97 \pm 0.07$ & $20.82 \pm 0.15$ & $22.02 \pm 0.10$ & 7 & $21.20 \pm 0.09$ & 6 & $21.15 \pm 0.08$ & 16 & $21.13 \pm 0.16$ & 9 & -- & -- & -- & -- \\
P030--17 & $>23.66$ & $>23.07$ & $23.51 \pm 0.26$ & $21.29 \pm 0.12$ & $20.84 \pm 0.19$ & $23.38 \pm 0.14$ & 17 & $20.94 \pm 0.01$ & 18 & $21.21 \pm 0.13$ & 17 & $21.00 \pm 0.15$ & 22 & -- & -- & -- & -- \\
P032--17 & $>23.69$ & $>23.49$ & $23.33 \pm 0.25$ & $21.23 \pm 0.08$ & $20.77 \pm 0.14$ & $23.47 \pm 0.14$ & 17 & $20.65 \pm 0.01$ & 18 & $20.53 \pm 0.08$ & 17 & $20.52 \pm 0.09$ & 22 & -- & -- & $20.00 \pm 0.16$ & 22 \\
J0224--3435 & -- & -- & $>24.23$ & -- & -- & $23.08 \pm 0.11$ & 4 & $21.78 \pm 0.08$ & 23 & $21.79 \pm 0.15$ & 23 & $21.73 \pm 0.16$ & 23 & $21.53 \pm 0.34$ & 23 & $21.11 \pm 0.26$ & 23 \\
P038--18 & $>23.69$ & $>23.49$ & $22.70 \pm 0.14$ & $20.57 \pm 0.05$ & $20.44 \pm 0.10$ & $21.53 \pm 0.03$ & 17 & $20.43 \pm 0.01$ & 18 & $20.47 \pm 0.09$ & 17 & $20.27 \pm 0.10$ & 22 & -- & -- & $20.29 \pm 0.28$ & 22 \\
P043--02 & $>23.51$ & $>23.80$ & $>23.80$ & $21.27 \pm 0.07$ & $20.69 \pm 0.12$ & $>24.41$ & 17 & $21.08 \pm 0.02$ & 18 & $21.15 \pm 0.12$ & 17 & $20.86 \pm 0.15$ & 22 & -- & -- & -- & -- \\
P050--18 & $>23.89$ & $>23.34$ & $>23.66$ & $21.19 \pm 0.08$ & $20.68 \pm 0.13$ & $24.05 \pm 0.33$ & 17 & $20.89 \pm 0.02$ & 18 & $20.79 \pm 0.08$ & 16 & $20.59 \pm 0.09$ & 22 & $20.21 \pm 0.16$ & 10 & $20.60 \pm 0.27$ & 22 \\
P065+01 & $>23.32$ & $>23.28$ & $22.72 \pm 0.21$ & $20.43 \pm 0.06$ & $20.09 \pm 0.08$ & -- & -- & $20.39 \pm 0.02$ & 18 & -- & -- & $19.74 \pm 0.05$ & 24 & -- & -- & $19.37 \pm 0.05$ & 13 \\
P072--07 & $>23.70$ & $>23.67$ & $23.11 \pm 0.21$ & $20.75 \pm 0.05$ & $20.92 \pm 0.14$ & -- & -- & $21.43 \pm 0.03$ & 18 & -- & -- & -- & -- & -- & -- & -- & -- \\
P076--10 & $>23.43$ & $>23.48$ & $>22.91$ & $20.73 \pm 0.05$ & $20.67 \pm 0.13$ & -- & -- & -- & -- & -- & -- & $20.37 \pm 0.17$ & 22 & -- & -- & $19.78 \pm 0.22$ & 22 \\
P119+02 & $>23.96$ & $>23.78$ & $22.59 \pm 0.14$ & $20.44 \pm 0.04$ & $20.27 \pm 0.08$ & -- & -- & $20.48 \pm 0.04$ & 18 & -- & -- & $19.96 \pm 0.08$ & 10 & $19.60 \pm 0.13$ & 10 & -- & -- \\
P124+12 & $>23.44$ & $>23.68$ & $23.26 \pm 0.28$ & $20.80 \pm 0.05$ & $21.20 \pm 0.15$ & -- & -- & $20.96 \pm 0.02$ & 18 & -- & -- & -- & -- & -- & -- & -- & -- \\
P127+26 & $>23.74$ & $>23.73$ & $>23.44$ & $20.72 \pm 0.06$ & $20.42 \pm 0.10$ & -- & -- & $20.38 \pm 0.02$ & 18 & $20.60 \pm 0.05$ & 25 & $20.53 \pm 0.06$ & 25 & $20.36 \pm 0.25$ & 25 & $20.08 \pm 0.17$ & 25 \\
P142--11 & $>23.66$ & $>23.57$ & $22.67 \pm 0.16$ & $20.64 \pm 0.05$ & $20.55 \pm 0.08$ & -- & -- & $20.63 \pm 0.04$ & 18 & -- & -- & -- & -- & -- & -- & $19.83 \pm 0.08$ & 14 \\
P148+69 & $>23.31$ & $>23.11$ & $22.98 \pm 0.28$ & $20.37 \pm 0.05$ & $20.32 \pm 0.15$ & -- & -- & $20.29 \pm 0.01$ & 18 & -- & -- & -- & -- & -- & -- & -- & -- \\
P151--16 & $>23.31$ & $>23.38$ & $>23.31$ & $21.24 \pm 0.08$ & $21.21 \pm 0.15$ & -- & -- & -- & -- & $21.11 \pm 0.22$ & 22 & $20.91 \pm 0.24$ & 22 & -- & -- & -- & -- \\
P156+38 & $>23.53$ & $23.64 \pm 0.30$ & $22.14 \pm 0.06$ & $19.84 \pm 0.02$ & $19.94 \pm 0.08$ & -- & -- & $20.32 \pm 0.01$ & 18 & -- & -- & $20.12 \pm 0.12$ & 20 & -- & -- & -- & -- \\
P158--14 & $>22.50$ & $>23.05$ & $22.25 \pm 0.25$ & $19.57 \pm 0.02$ & $19.28 \pm 0.04$ & -- & -- & -- & -- & $19.28 \pm 0.07$ & 25 & $19.25 \pm 0.08$ & 25 & $19.15 \pm 0.10$ & 25 & $18.68 \pm 0.09$ & 25 \\
P169+58 & $>22.94$ & $>23.51$ & $22.65 \pm 0.17$ & $20.46 \pm 0.06$ & $20.82 \pm 0.20$ & -- & -- & $20.74 \pm 0.02$ & 18 & -- & -- & -- & -- & -- & -- & -- & -- \\
P170+20 & $>23.74$ & $22.95 \pm 0.19$ & $22.43 \pm 0.29$ & $20.38 \pm 0.03$ & $20.17 \pm 0.06$ & -- & -- & $20.31 \pm 0.02$ & 18 & -- & -- & $19.85 \pm 0.11$ & 20 & -- & -- & -- & -- \\
P173--06 & $>23.70$ & $>23.91$ & $22.89 \pm 0.20$ & $20.43 \pm 0.05$ & $20.86 \pm 0.12$ & -- & -- & $20.78 \pm 0.02$ & 18 & $20.76 \pm 0.20$ & 22 & $20.30 \pm 0.13$ & 22 & $20.12 \pm 0.15$ & 22 & $20.08 \pm 0.25$ & 22 \\
P173+48 & $>24.24$ & $>23.82$ & $>23.78$ & $21.59 \pm 0.10$ & $21.81 \pm 0.29$ & -- & -- & $21.34 \pm 0.03$ & 18 & -- & -- & -- & -- & -- & -- & -- & -- \\
P175+71 & $>23.40$ & $>23.24$ & $23.19 \pm 0.31$ & $20.28 \pm 0.04$ & $20.57 \pm 0.12$ & -- & -- & $20.58 \pm 0.02$ & 18 & -- & -- & -- & -- & -- & -- & -- & -- \\
J1152+0055 & -- & -- & $>24.37$ & -- & -- & $>24.39$ & 2 & $22.12 \pm 0.14$ & 23 & $21.57 \pm 0.17$ & 23 & $21.64 \pm 0.23$ & 23 & $21.53 \pm 0.35$ & 23 & $21.29 \pm 0.30$ & 23 \\
P178+28 & $>23.66$ & $>23.91$ & $21.95 \pm 0.08$ & $19.83 \pm 0.03$ & $19.82 \pm 0.06$ & -- & -- & $20.01 \pm 0.02$ & 18 & -- & -- & $19.63 \pm 0.09$ & 20 & -- & -- & -- & -- \\
P182+53 & $>24.11$ & $>23.95$ & $>23.62$ & $21.16 \pm 0.07$ & $20.89 \pm 0.13$ & -- & -- & $20.89 \pm 0.03$ & 18 & -- & -- & -- & -- & -- & -- & -- & -- \\
P193--02 & $>24.04$ & $>24.03$ & $23.23 \pm 0.28$ & $21.14 \pm 0.07$ & $21.15 \pm 0.12$ & -- & -- & $21.07 \pm 0.03$ & 18 & $20.88 \pm 0.19$ & 21 & $20.63 \pm 0.17$ & 21 & $20.68 \pm 0.20$ & 21 & -- & -- \\
P196+15 & $>23.77$ & $>23.66$ & $23.43 \pm 0.28$ & $20.58 \pm 0.06$ & $20.54 \pm 0.12$ & -- & -- & $20.86 \pm 0.02$ & 18 & $20.68 \pm 0.14$ & 21 & $20.56 \pm 0.17$ & 21 & -- & -- & $19.69 \pm 0.14$ & 21 \\
P197+45 & $>23.99$ & $>23.88$ & $23.29 \pm 0.25$ & $20.88 \pm 0.05$ & $20.81 \pm 0.12$ & -- & -- & $20.65 \pm 0.02$ & 18 & -- & -- & $20.29 \pm 0.18$ & 20 & -- & -- & -- & -- \\
P207+37 & $>23.92$ & $>23.45$ & $23.29 \pm 0.27$ & $20.83 \pm 0.05$ & $20.57 \pm 0.09$ & -- & -- & $20.72 \pm 0.02$ & 18 & -- & -- & $20.37 \pm 0.17$ & 20 & -- & -- & -- & -- \\
P207--21 & $>23.42$ & $>23.58$ & $>23.56$ & $20.62 \pm 0.05$ & $20.72 \pm 0.16$ & -- & -- & -- & -- & $20.48 \pm 0.13$ & 22 & $20.54 \pm 0.19$ & 22 & -- & -- & $20.01 \pm 0.23$ & 22 \\
P209--08 & $>23.94$ & $>23.55$ & $>23.55$ & $21.10 \pm 0.06$ & $21.41 \pm 0.17$ & -- & -- & $21.50 \pm 0.07$ & 18 & -- & -- & -- & -- & -- & -- & -- & -- \\
P215+26 & $>23.83$ & $>23.83$ & $>23.93$ & $21.12 \pm 0.06$ & $20.43 \pm 0.08$ & -- & -- & $20.60 \pm 0.03$ & 18 & -- & -- & $20.40 \pm 0.16$ & 20 & -- & -- & -- & -- \\
P218+28 & $>23.88$ & $>23.55$ & $23.29 \pm 0.33$ & $20.81 \pm 0.06$ & $20.63 \pm 0.13$ & -- & -- & $20.60 \pm 0.02$ & 18 & -- & -- & $20.13 \pm 0.17$ & 20 & -- & -- & -- & -- \\
P218+04 & $>23.85$ & $>23.86$ & $23.30 \pm 0.31$ & $20.98 \pm 0.06$ & $20.57 \pm 0.10$ & -- & -- & $20.88 \pm 0.03$ & 18 & $20.98 \pm 0.21$ & 21 & $21.04 \pm 0.33$ & 21 & -- & -- & -- & -- \\
P224+10 & $>24.09$ & $23.71 \pm 0.31$ & $21.87 \pm 0.08$ & $19.88 \pm 0.02$ & $19.80 \pm 0.04$ & -- & -- & $19.90 \pm 0.01$ & 18 & $19.82 \pm 0.07$ & 21 & $19.23 \pm 0.06$ & 20 & $19.02 \pm 0.06$ & 21 & $18.93 \pm 0.05$ & 21 \\
P228+01 & $>23.83$ & $>23.67$ & $23.26 \pm 0.25$ & $21.20 \pm 0.06$ & $21.10 \pm 0.18$ & -- & -- & $21.79 \pm 0.08$ & 18 & -- & -- & $21.94 \pm 0.13$ & 12 & -- & -- & -- & -- \\
P261+37 & $>23.64$ & $>23.78$ & $23.36 \pm 0.18$ & $20.87 \pm 0.06$ & $20.61 \pm 0.09$ & -- & -- & $21.12 \pm 0.03$ & 18 & -- & -- & -- & -- & -- & -- & -- & -- \\
P265+41 & $>23.57$ & $>23.95$ & $>24.25$ & $20.83 \pm 0.06$ & $21.17 \pm 0.15$ & -- & -- & $20.45 \pm 0.01$ & 18 & -- & -- & $20.30 \pm 0.16$ & 20 & -- & -- & -- & -- \\
P271+49 & $>23.75$ & $>23.89$ & $22.84 \pm 0.15$ & $20.19 \pm 0.04$ & $20.56 \pm 0.10$ & -- & -- & $20.59 \pm 0.02$ & 18 & -- & -- & $20.26 \pm 0.18$ & 20 & -- & -- & -- & -- \\
P281+53 & $>23.77$ & $>23.64$ & $23.88 \pm 0.35$ & $20.17 \pm 0.03$ & $20.25 \pm 0.10$ & -- & -- & $19.90 \pm 0.01$ & 18 & -- & -- & $19.72 \pm 0.11$ & 20 & -- & -- & -- & -- \\
P288+63 & $>23.81$ & $>23.40$ & $23.27 \pm 0.27$ & $20.68 \pm 0.06$ & $20.76 \pm 0.15$ & -- & -- & $20.59 \pm 0.03$ & 18 & -- & -- & -- & -- & -- & -- & -- & -- \\
P306-04 & $>23.76$ & $>23.54$ & $22.63 \pm 0.13$ & $20.53 \pm 0.05$ & $20.64 \pm 0.14$ & -- & -- & $20.79 \pm 0.04$ & 18 & -- & -- & -- & -- & -- & -- & -- & -- \\
P307-05 & $nan \pm 0.31$ & $>23.77$ & $>23.85$ & $20.53 \pm 0.06$ & $21.01 \pm 0.17$ & -- & -- & $20.90 \pm 0.03$ & 18 & -- & -- & $20.75 \pm 0.23$ & 22 & -- & -- & -- & -- \\
J2211--3206 & -- & -- & $>24.26$ & -- & -- & $>23.71$ & 19 & $19.91 \pm 0.02$ & 23 & $19.85 \pm 0.03$ & 23 & $19.62 \pm 0.03$ & 23 & $19.38 \pm 0.05$ & 23 & $19.00 \pm 0.03$ & 23 \\
P334--05 & $>23.77$ & $>23.67$ & $>23.89$ & $21.53 \pm 0.11$ & $21.04 \pm 0.19$ & -- & -- & $21.38 \pm 0.04$ & 18 & -- & -- & $20.71 \pm 0.23$ & 22 & -- & -- & -- & -- \\
J2227--3323 & -- & -- & $>24.50$ & -- & -- & $24.09 \pm 0.36$ & 5 & $21.91 \pm 0.11$ & 23 & $21.43 \pm 0.15$ & 23 & $21.62 \pm 0.20$ & 23 & $21.28 \pm 0.22$ & 23 & $21.25 \pm 0.24$ & 23 \\
J2315--2856 & $>23.70$ & $>23.32$ & $23.66 \pm 0.27$ & $21.26 \pm 0.12$ & $21.15 \pm 0.21$ & $22.29 \pm 0.09$ & 5 & $21.20 \pm 0.06$ & 23 & $21.23 \pm 0.14$ & 23 & $21.09 \pm 0.12$ & 23 & $20.95 \pm 0.15$ & 23 & $20.53 \pm 0.11$ & 23 \\
J2318--3029 & $>23.24$ & $>22.73$ & $>24.19$ & $20.66 \pm 0.18$ & $20.08 \pm 0.14$ & $22.33 \pm 0.09$ & 1 & $20.58 \pm 0.04$ & 23 & $20.58 \pm 0.08$ & 23 & $20.20 \pm 0.06$ & 23 & $20.00 \pm 0.06$ & 23 & $19.69 \pm 0.06$ & 23
\enddata
\tablerefs{(1) EFOSC2 2013-09-28, (2) EFOSC2 2014-03-03, (3) EFOSC2 2014-07-24, (4) EFOSC2 2014-07-27, (5) EFOSC2 2015-07-22, (6) EFOSC2 2016-09-13, (7) EFOSC2 2016-09-14, (8) EFOSC2 2018-06-26, (9) GROND 2016-09-22, (10) GROND 2017-01-01, (11) SOFI 2017-05-29, (12) SOFI 2018-06-26, (13) SOFI 2019-12-13, (14) SOFI 2019-12-14, (15) SOFI 2020-11-20, (16) Retrocam 2016-09-20, (17) DES DR2, (18) DELS DR9, (19) KiDS, (20) UHS, (21) UKIDSS, (22) VHS, (23) VIKING, (24) \cite{bischetti2022}, (25) \cite{ross2020}}\tablecomments{
The PS1 magnitudes are \textit{dereddened}.
Reddened magnitudes can be obtained by adding
$\lambda_f \times E(B-V)$ with $\lambda_f = (3.172, 2.271, 1.682, 1.322, 1.087)$  for (\gps,\rps, \ips, \zps, \yps),
see \cite{schlafly2011}. For the VIKING quasars, the magntides reported in the \ips\ column are actually $i$ magnitudes from the KiDS survey.
Limits are reported at $3\sigma$.
}

\end{deluxetable*}

%% file: tab4.tex
\begin{splitdeluxetable*}{lcccccccBcccccccc}

    \centerwidetable
\movetabledown=5mm
    \tablewidth{0pt}
    \tabletypesize{\footnotesize}

\tablecaption{Properties of the quasars detected in public radio surveys, sorted by right ascension. \label{tab:radio-info}}
\tablehead{\colhead{Quasar} & \colhead{$z$} & \colhead{$\mathrm{LoTTS}_{144\,\mathrm{MHz}}$} & \colhead{LoTTS ref} & \colhead{$\mathrm{RACS}_{888\,\mathrm{MHz}}$} & \colhead{RACS ref} & \colhead{$\mathrm{FIRST}_{1.4\,\mathrm{GHz}}$} & \colhead{FIRST ref} & \colhead{$\mathrm{VLASS}_{3\,\mathrm{GHz}}$ Ep 1} & \colhead{$\mathrm{VLASS}_{3\,\mathrm{GHz}}$ Ep 2} & \colhead{VLASS ref} & \colhead{\slopelofarfirst} & \colhead{\sloperacsfirst} & \colhead{\sloperacsvlass} & \colhead{\slopefirstvlass} & \colhead{$R_{2500}$}\\ \colhead{ } & \colhead{ } & \colhead{$\mu$Jy} & \colhead{ } & \colhead{$\mu$Jy} & \colhead{ } & \colhead{$\mu$Jy} & \colhead{ } & \colhead{$\mu$Jy} & \colhead{$\mu$Jy} & \colhead{*} & \colhead{ } & \colhead{ } & \colhead{ } & \colhead{ } & \colhead{ }}
\startdata
P050--18 & 6.09 & -- & -- & $1905\pm350$ & DR1 & -- & -- & $749\pm183$ & $738\pm153$ & 1.2/2.2 & -- & -- & $-0.77\pm0.25$ & -- & $156\pm31$ \\
P173+48 & 6.233 & $4673\pm323$ & DR2 & -- & -- & $3226\pm135$ & v2014dec17 & $2360\pm140$ & $1910\pm150$ & 1.2/2.2 & $-0.16\pm0.04$ & -- & -- & $-0.41\pm0.10/-0.69\pm0.12$ & $734-1067$ \\
P182+53 & 5.99 & $240\pm58$ & forced & -- & -- & $<438$ & forced & $<349$ & $<419$ & 1.1/2.1 & $<0.26$ & -- & -- & -- & $18\pm4$ \\
P193--02\tablenotemark{a} & 5.8 & -- & -- & $4048\pm660$ & DR1 & $509\pm151$ & forced & $<496$ & $<438$ & 1.2/2.2 & -- & $-4.55\pm0.74$ & $<-1.83$ & -- & $108-1315$ \\
P207+37 & 5.69 & $485\pm151$ & DR2 & $<615$ & forced & $<440$ & forced & $<370$ & $<409$ & 1.1/2.1 & $<-0.04$ & -- & -- & -- & $27\pm9$ \\
\enddata
\tablecomments{A reference `forced' means that there was no entry in the catalog and we have measured the flux from the images. For VLASS we always measure the peak flux densities directly from the images.}\tablenotetext{*}{The peak flux densities for the VLASS1.1 epoch are known to be systematically low by $\approx 15\%$ and by $\approx8\%$  for the subsequent epochs (1.2/2.1/2.2); see   \url{https://science.nrao.edu/vlass/data-access/vlass-epoch-1-quick-look-users-guide}. We have taken into account these factors in the reported values.}
\tablenotetext{a}{The radio-loudness of P193--02 is very uncertain because its strong RACS detection could have been an AGN flare, see discussion in Section~\ref{sec:radio}.}

\end{splitdeluxetable*}